\documentclass[aps]{revtex4}
\usepackage{amsfonts}
\usepackage{amsmath}
\usepackage{amssymb,epsf}
\usepackage{color}

\begin{document}

\title{Higher dimensional dyonic black holes}
\author{S. H. Hendi$^{1,2}$\footnote{
email address: hendi@shirazu.ac.ir}, N. Riazi$^{3}$\footnote{
email address: n$_{-}$riazi@sbu.ac.ir}, S.
Panahiyan$^{1,3,4}$\footnote{
email address: shahram.panahiyan@uni-jena.de} and B. Eslam Panah$^{1,2,5}$%
\footnote{ email address: behzad.eslampanah@gmail.com }}
\affiliation{$^1$ Physics Department and Biruni Observatory,
College of Sciences, Shiraz
University, Shiraz 71454, Iran\\
$^2$ Research Institute for Astronomy and Astrophysics of Maragha (RIAAM),
P.O. Box 55134-441, Maragha, Iran\\
$^3$ Physics Department, Shahid Beheshti University, Tehran 19839, Iran\\
$^4$ Helmholtz-Institut Jena, Fr\"{o}belstieg 3, Jena D-07743, Germany\\
$^5$ ICRANet, Piazza della Repubblica 10, I-65122 Pescara, Italy}

\begin{abstract}
The paper at hand presents a novel class of dyonic black holes in higher
dimensions through a new proposal for the electromagnetic field tensor. The
black hole solutions are extracted analytically and their
geometrical/physical properties are studied. In addition, the details
regarding thermodynamical structure and phase transition behavior of the
solutions for $4$ different cases are investigated: i) general case, ii)
constant electric field, iii) constant magnetic field, and iv) constant
electric and magnetic fields. It will be shown that depending on the picture
under consideration, the thermodynamical properties are modified. To have
better picture regarding the phase transitions, the concept of the extended
phase space is employed. It will be shown that in the absence of the
electric field, magnetic black holes present van der Waals like phase
transition. Furthermore, it will be highlighted that for super magnetized
black holes, no phase transition exists.
\end{abstract}

\maketitle

\section{Introduction}

Gravity in higher dimensions has been of interest for several decades. This
is due to fact that specific range of the advanced physics theories requires
existence of the higher dimensions in order to address different issues in
the nature. To name a few, one can point out: I) String, superstring
theories and in general M-theory, which have been the most celebrated
theories towards unification of the all fundamental forces in nature \cite%
{Antoniadis}. II) Braneworlds theories which have been employed to address
fundamental issues of gravity such as hierarchy problem \cite{Randall}. III)
Kaluza-Klein compactification which was the pioneering proposal regarding
unification of the gravity and electromagnetism \cite{Kaluza,Klein}. Through
all of the mentioned thoeries, the necessity of existence of the higher
dimensionality were highlighted and employed.

In the context of black holes, it was shown that although the laws of black
hole mechanic are universal, the properties of black hole are dimension
dependent \cite{Thermodynamics}. Reissner-Nordstrom (RN) black holes in the
context of string theory plays an important role in understanding the black
hole entropy near extremal limits \cite{Strominger}, hence and in the
context of this theory, Hawking temperature, radiation rate and entropy for
these black holes have been studied and it was proposed that quantum
evolution of black hole does not lead to information loss \cite{Callan}.
Hereupon, the study of black holes in higher dimensions has attracted many
authors. For example; the generalizations of Schwarzschild and Kerr black
holes to arbitrary extra dimensions have been investigated in refs. \cite%
{Tangherlini} and \cite{Myers}, respectively. The existence of black rings
and Saturns in higher dimensions have been studied \cite%
{Emparan,Elvang,Iguchi}. The thermodynamics and stability of higher
dimensional Kerr-anti de Sitter black hole has been addressed in ref. \cite%
{Carter}.

In ref. \cite{Kim}, the ultraviolet divergent structures of the matter field
in a higher dimensional RN black hole has been studied and the contributions
to Bekenstein-Hawking entropy by using the Pauli-Villars regularization
method was addressed. Gravitation with superposed Gauss--Bonnet terms and
black object solutions in higher dimensions have been obtained in ref. \cite%
{GB}. Uniqueness and non-uniqueness of static (un)charged black holes and
black p-branes in higher dimensions have been surveyed \cite{Gibbons}.
Topology of black holes' event horizons in higher dimensions has been
investigated in ref. \cite{Helfgott}. Hawking emission of gravitons and
generalization of Hawking's black hole topology theorem to higher dimensions
have been obtained, respectively in refs. \cite{Cardoso} and \cite{Galloway}%
. Quasinormal modes of Schwarzschild \cite{CardosoII} and Kerr \cite{Kao}
black holes in higher dimensions have been studied. In addition, the
production of higher-dimensional black holes in future colliders becomes a
conceivable possibility in scenarios involving large extra dimensions and
TeV-scale gravity \cite{Cavaglia}. In addition, as mathematical objects,
black hole spacetimes are among the most important Lorentzian Ricci-flat
manifolds in any dimension \cite{EmparanII}.

Another motivation for considering higher dimensions is related to the
AdS/CFT correspondence which relates the properties of a black hole in
d-dimensions with those of a quantum field theory in (d-1)-dimensions \cite%
{Aharony}. In other words, we can extract the properties of a complex system
in quantum field theory by using the properties of black holes in one higher
dimension. Among the different set ups for AdS/CFT studies, the ones
including a magnetic field has been of special interests. Historically
speaking, Hartnoll and Kovtun in their pioneering work incorporated a
background magnetic field and studied low frequency charge transport and
Hall conductivity \cite{Hartnoll}. The magnetic field was included by
considering a type of black holes known as \textit{dyonic black holes}.
Later, it was shown that large dyonic black holes in anti-de Sitter
spacetime are dual to stationary solutions of the equations of relativistic
magnetohydrodynamics on the conformal boundary of AdS \cite{Caldarelli}. In
addition, the dyonic black holes was employed to induce the effects of
external magnetic field on superconductors. It was shown that the size of
condensate for the superconductor is magnetic field dependent in a manner
which is a reminiscent of the Meissner effect \cite{Albash}. Furthermore,
the holographical properties of the dyonic dilatonic black branes including
transport coefficients, Hall conductance, DC longitudinal conductivity and
response were investigated in ref. \cite{Goldstein}. So far, a large number
of publications was dedicated to study systems including dyonic black holes
in different contexts (for a very incomplete list, we refer the reader to
refs. \cite%
{Dyonic1,Dyonic2,Dyonic3,Dyonic4,Dyonic5,Dyonic6,Dyonic7,Dyonic8,Dyonic9,Dyonic10,Dyonic11,Dyonic12,Dyonic13,Dyonic14,Dyonic15,Dyonic16,Dyonic17, Dyonic18,Dyonic19,Dyonic20,Dyonic21,Dyonic22,Dyonic23,Dyonic24}%
).

In this paper, we introduce a novel approach for constructing
electrically-magnetically charged black holes, or simply put dyonic black
hole holes, in higher dimension. Our main motivation is to propose a
simplified method for constructing higher dimensional dyonic black holes in
a manner that magnetic and electric parts of it are stand alone properties.
We expand our study to thermodynamical properties of the black holes in
order to understand the physical and geometrical properties of the solutions
in details. We consider four distinctive cases which correspond to four
different scenarios in the context of AdS/CFT correspondence; I) \textit{%
General case}: in which no specification is given about electric and
magnetic field. II) \textit{Constant electric field}: which corresponds to
immersing the magnetically charged black holes in a finite electric field.
III) \textit{Constant magnetic field}: which corresponds to considering the
electrically charged black holes in the presence of finite external magnetic
field. IV) \textit{Constant electric and magnetic fields}: which corresponds
to immersing the black holes in a field consisting finite electric and
magnetic fields. We will show how the magnetization, electrification and
dimensionality of the solutions affect physical properties of the black
holes including the behaviors of temperature, enthalpy and heat capacity. In
addition, we will explore the possibility of the van der Waals like phase
transition for the four mentioned cases and investigate the effects of
magnetization, electrification and dimensionality on van der Waals like
critical points and phase transition. Among different benefits of our
proposal, we can point out two important ones: I) The set up is easy to
understand and could be employed without going into trouble of introducing
complex system of the equations. II) The set up provides the possibility of
including higher dimensional gravities such as Lovelock gravity with
simplicity. We intend to provide the possibility of studying the effects of
magnetism on holographical systems in higher dimensions (with or without
higher dimensional gravity theories). The set up also could be employed to
study the effects of non-finite magnetic/electric field on holographical
systems.

The structure of the paper is as follows: first, the action and field
equations are introduced. The metric function is obtained and, geometrical
properties and conditions regarding the existence of black holes are
investigated. In section III, thermodynamical quantities of interest in this
paper are introduced and obtained. Sections IV-VII are, respectively,
dedicated to investigation of four different cases including: General case,
constant electric field, constant magnetic field and, constant electric and
magnetic fields. The paper is concluded with some closing remarks.

\section{Basic Equations}

The main goal of this paper is construction of the novel dyonic black holes
in higher dimensions. Our motivation comes from interesting properties of
the dyonic black holes specially in holographical aspects, DC conductivity
string theory, etc. Here, we introduce higher dimensional dyonic black holes
which could be employed to understand the DC effects in higher dimensions.
Furthermore, we are providing the possibility of studying higher dimensional
theories of gravity such as Lovelock gravity in the presence of dyonic
configuration as well.

Dyonic black holes enjoy existence of magnetic charge as well as electric
charge in their structures. It is worthwhile to mention that the set up
which is going to be introduced here, provides the possibility of having
magnetic field for black holes without introduction of the electric field.
For simplicity, we consider Einstein Lagrangian in the presence of the
cosmological constant as the gravitational sector of the action. As for the
matter field, we simply consider the Maxwell Lagrangian with modified vector
potential. Therefore, the $d$-dimensional action will be given by

\begin{equation}
\mathcal{I}=-\frac{1}{16\pi G_{d}}\int d^{d}x\sqrt{-g}\left[ \mathcal{R}%
-2\Lambda -F^{\mu \nu }F_{\mu \nu }\right] ,  \label{Action}
\end{equation}%
in which, $\mathcal{R}$ is the Ricci scalar, $\Lambda$ refers to the
cosmological constant and $F_{\mu \nu }$ is the electromagnetic field
tensor. It is worthwhile to mention the possibility of generalization of
this action to include higher dimensional theories of gravity, scalar-tensor
field theories and nonlinear electromagnetic field. The magnetic charge,
hence the dyonic property lies within the structure of electromagnetic
tensor. The $d$-dimensional metric with topological boundary of $t=cte$ and $%
r=cte$, is given by
\begin{equation}
ds^{2}=-f(r)dt^{2}+\frac{dr^{2}}{f(r)}+r^{2}d\Omega _{k}^{2},  \label{Metric}
\end{equation}%
in which $d\Omega _{k}^{2}$ is the line element of a $(d-2)$-dimensional
hypersurface with the constant curvature $(d-2)(d-3)k$ and volume $V_{d-2}$
with the following explicit form
\begin{equation}
d\Omega _{k}^{2}=\left\{
\begin{array}{cc}
d\theta _{1}^{2}+\sum\limits_{i=2}^{d-2}\prod\limits_{j=1}^{i-1}\sin
^{2}\theta _{j}d\theta _{i}^{2} & k=1 \\
d\theta _{1}^{2}+\sinh ^{2}\theta _{1}d\theta _{2}^{2}+\sinh ^{2}\theta
_{1}\sum\limits_{i=3}^{d-2}\prod\limits_{j=2}^{i-1}\sin ^{2}\theta
_{j}d\theta _{i}^{2} & k=-1 \\
\sum\limits_{i=1}^{d-2}d\phi _{i}^{2} & k=0%
\end{array}%
\right. .  \label{dOmega}
\end{equation}

In order to have consistent field equations with magnetic charge included,
we modify the electromagnetic tensor with the following non-zero components

\begin{equation}
F_{tr}=-F_{rt}=\frac{q_{E}}{r^{d-2}}\text{ \ \ \ \ \ \ \& \ \ \ \ \ \ }%
F_{\theta \phi }=-F_{\phi \theta }=\frac{q_{M}}{r^{d-4}}\Upsilon(\theta)
\label{Ftr}
\end{equation}%
in which $q_{E}$ and $q_{M}$ are respectively, electric and magnetic
charges, and

\begin{equation}
\Upsilon (\theta )=\left\{
\begin{array}{cc}
\sin \theta & k=1 \\
\theta & k=0 \\
\sinh \theta & k=-1%
\end{array}%
\right. .
\end{equation}

The $F_{tr}$ is representing the electric part of the electromagnetic field
while $F_{\theta \phi }$ is related to the magnetic part. There are several
issues that must be pointed out; first of all, the electromagnetic field
tensor has been generated by both electric and magnetic charges separately.
Therefore, it is possible to cancel out the electric part by setting $%
q_{E}=0 $ and have magnetic black hole solutions. Second, we have restricted
the magnetic field to one direction which is a common practice for the
magnetic charges. Finally, except for $4$-dimensional case which has
constant magnetic field, for large values of the $r$, magnetic field will
vanish similar to the electric field which is physically expected. One of
the important properties of the proposed electromagnetic tensor is that
magnetic field is a \textit{stand alone property}. In other words, even in
the absence of electric charge, one can construct magnetic black holes
without resorting to complex field equations.

Using the variational principle, it is a matter of calculation to reach the
following field equation

\begin{equation}
e_{\mu \nu }\equiv G_{\mu \nu }+\Lambda g_{\mu \nu }-\left[ 2F_{\mu \lambda
}F_{\nu }^{\lambda }-\frac{1}{2}g_{\mu \nu }F^{\sigma \rho }F_{\sigma \rho }%
\right] =0,  \label{field}
\end{equation}%
which by considering metric (\ref{Metric}) and non-zero components of the
electromagnetic field tensor (\ref{Ftr}), one finds

\begin{eqnarray*}
e_{tt} &=&e_{rr}=\left( d-2\right) \left( d-3\right) \left[ f\left( r\right)
-k\right] r^{4d-10}+\left[ 2\Lambda r+\left( d-2\right) \left( \frac{%
df\left( r\right) }{dr}\right) \right] r^{4d-9}+2\left[ q_{E}^{2}+q_{M}^{2}%
\right] r^{2d-4} =0, \\
e_{ii} &=&\left( d-3\right) \left( d-4\right) \left[ f\left( r\right) -k%
\right] r^{4d-10}+\left[ 2\Lambda r+2\left( d-3\right) \left( \frac{df\left(
r\right) }{dr}\right) +\left( \frac{d^{2}f\left( r\right) }{dr^{2}}\right) r%
\right] r^{4d-9}-2\left[ q_{E}^{2}+q_{M}^{2}\right] r^{2d-4} =0.
\end{eqnarray*}

Solving these two equations with respect to metric function, one obtains $%
f\left(r\right)$ as

\begin{equation}
f\left( r\right) =k-\frac{m}{r^{d-3}}-\frac{2\Lambda r^{2}}{\left(
d-1\right) \left( d-2\right) }+\frac{2\left( q_{E}^{2}+q_{M}^{2}\right) }{%
\left( d-2\right) \left( d-3\right) r^{2d-6}},  \label{metric function}
\end{equation}%
where $m$ is geometrical mass related to total mass of the black hole. By
setting $d=4$, metric function yields
\begin{equation}
\psi (r)=k-\frac{\Lambda r^{2}}{3}-\frac{m}{r}+\frac{q_{E}^{2}+q_{M}^{2}}{%
r^{2}},
\end{equation}
which was previously obtained in ref. \cite{dyonicmassless}. This shows that
our proposal includes other types of dyonic black holes as well.

Existence of the black hole solutions depends on satisfaction of specific
conditions simultaneously: i) existence of the singularity, and ii) presence
of at least one horizon which covers the singularity and is known as event
horizon.

The existence of singularity could be determined by divergencies of the
curvature scalars. One of the well known curvature scalars is the
Kretschmann scalar. It is a matter of calculation to find the Kretschmann
for these solutions in the following form

\begin{equation}
R_{\alpha \beta \gamma \delta }R^{\alpha \beta \gamma \delta }=\left( \frac{%
d^{2}f\left( r\right) }{dr^{2}}\right) +\frac{2\left( d-2\right) }{r^{2}}%
\left( \frac{df\left( r\right) }{dr}\right) ^{2}+\frac{2\left( d-2\right)
\left( d-3\right) }{r^{4}}\left( f\left( r\right) -k\right) ^{2},
\end{equation}%
where

\begin{eqnarray*}
\left( \frac{d^{2}f\left( r\right) }{dr^{2}}\right)  &=&\frac{4\left(
2d-5\right) }{\left( d-2\right) r^{2d-4}}\left( q_{E}^{2}+q_{M}^{2}\right) -%
\frac{\left( d-2\right) \left( d-3\right) }{r^{d-1}}m-\frac{4}{\left(
d-1\right) \left( d-2\right) }\Lambda , \\
\left( \frac{df\left( r\right) }{dr}\right)  &=&\frac{\left( d-3\right) }{%
r^{d-2}}m-\frac{4r}{\left( d-1\right) \left( d-2\right) }\Lambda -\frac{4}{%
\left( d-2\right) r^{2d-5}}\left( q_{E}^{2}+q_{M}^{2}\right) .
\end{eqnarray*}

The Kretschmann has the following limit

\begin{equation}
\lim_{r\longrightarrow 0}R_{\alpha \beta \gamma \delta }R^{\alpha \beta
\gamma \delta }\longrightarrow \infty .  \label{inf}
\end{equation}

Eq. \ref{inf} confirms the existence of a curvature singularity at $r=0$. By
series expansion of this curvature scalar for small values of $r$, one can
find the following relation

\begin{equation*}
\lim_{r\longrightarrow 0}R_{\alpha \beta \gamma \delta }R^{\alpha \beta
\gamma \delta }\propto \left(
a_{1}q_{E}^{4}+a_{2}q_{E}^{2}q_{M}^{2}+a_{3}q_{M}^{4}\right) r^{-4\left(
d-2\right) },
\end{equation*}%
in which $a_{i}$ are dimension dependent coefficients. It is interesting to
note that the singularity is affected by both electric and magnetic fields
with same order of magnitude. In the absence of electric field, the dominant
term on singular behavior is the magnetic charge. This highlights the
contribution of the magnetic part. It is worthwhile to mention that power of
the singularity is stronger in higher dimensions and the divergency is
reached faster compared to lower dimensions.

Existence of the cosmological constant in the solutions, provides specific
complexity which prevents us to extract the root(s) of the metric function,
hence event horizon, analytically. In the absence of the cosmological
constant ($\Lambda =0$), the roots of metric function are obtained as

\begin{equation}
r\left( f\left( r\right) =0\right) =\left[ \frac{m\left( d-2\right) \left(
d-3\right) \pm \sqrt{m^{2}\left( d-2\right) ^{2}\left( d-3\right)
^{2}-8k\left( d-2\right) \left( d-3\right) \left( q_{E}^{2}+q_{M}^{2}\right)
}}{4\left( q_{E}^{2}+q_{M}^{2}\right) }\right] ^{\frac{-1}{d-3}},
\label{rootfr}
\end{equation}%
which shows that under certain conditions, two distinct roots for the metric
function may exists. The first condition comes from positivity of the
expression under square root function. In other words, in order to have real
roots for the metric function, the square root function must be positive
valued which results into the following condition

\begin{equation}
m^{2}\geq \frac{8k\left( q_{E}^{2}+q_{M}^{2}\right) }{\left( d-2\right)
\left( d-3\right) }.  \label{con1}
\end{equation}

This condition has several points which must be highlighted; considering
that $m$, $q_{E}$ and $q_{M}$ are positive values, the mentioned condition
is valid only for spherical case, $k=1$. In other words, for the horizon
flat ($k=0$) and hyperbolic horizon ($k=-1$), the square root function is
always positive valued and mentioned condition is satisfied irrespective of
choices for $m$, $q_{E}$ and $q_{M}$. If this condition is violated, no real
valued root exists for the metric function which indicates that such
solutions is a naked singularity. Therefore, it is safe to state that naked
singularity only exists for the spherical cases while for horizon flat and
hyperbolic horizon, obtained solutions definitely enjoy at least one root,
hence event horizon in their structure. It is interesting to note that for
Ricci flat solutions ($k=0$), the inner horizon goes to $r=0$, and
therefore, the singularity is null.

The coupling between topological factor and the electric and magnetic
charges is another important issue. Here, we see that the presence of
magnetic charge affected the condition regarding the existence of real
valued roots. On the other hand, we see that the presence of magnetic charge
has its own effects on the position of root as well. In the absence of the
electric charge, the magnetic charge for spherical case upholds the
mentioned condition. Returning to obtained root for metric function (\ref%
{rootfr}), we see that it is possible to have two real positive valued roots
provided that the following inequality holds for negative branch

\begin{equation}
m\left( d-2\right) \left( d-3\right) -\sqrt{m^{2}\left( d-2\right)
^{2}\left( d-3\right) ^{2}-8k\left( d-2\right) \left( d-3\right) \left(
q_{E}^{2}+q_{M}^{2}\right) }>0.  \label{con2}
\end{equation}

Now, considering that for $k=-1,0$, the square root function is positive
valued, one can state that for these two cases, the mentioned condition (\ref%
{con2}) is violated. Therefore, it is safe to conclude that for $k=-1,0$,
only one root exists. On the contrary, for spherical horizon, it is possible
to have three cases: i) violation of the condition (\ref{con1}) which
results in naked singularity. ii) if $m^{2}=\frac{8k\left(
q_{E}^{2}+q_{M}^{2}\right) }{\left( d-2\right) \left( d-3\right) }$, then
positive and negative branches of the obtained roots coincide which results
in extreme black hole solutions (existence of one root). iii) satisfaction
of mentioned conditions ((\ref{con1}) and (\ref{con2})) which results in
existence of two roots for metric function.

\begin{figure}[tbp]
$%
\begin{array}{c}
\epsfxsize=7cm \epsffile{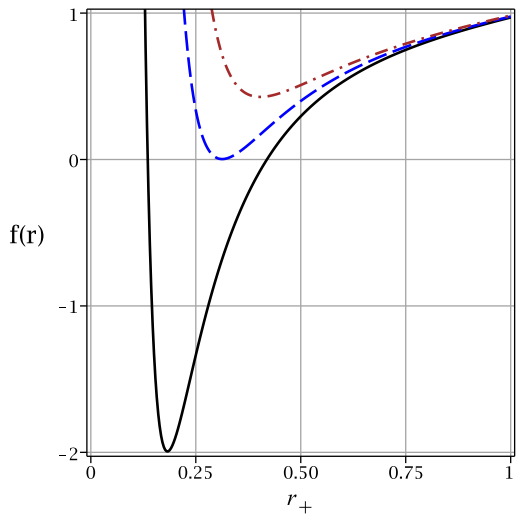}%
\end{array}
$%
\caption{$f(r)$ versus $r$ for $d=5$, $\Lambda =-1$, $m=0.2$, $q_{E}=0.1$, $%
k=1$; $q_{M}=0$ (continuous line), $q_{M}=0.14$ (dashed line) and $q_{M}=0.2$
(dashed-dotted line).}
\label{Figfr}
\end{figure}

Since it was not possible to obtain the root of metric function in the
presence of cosmological constant, we have employed numerical method to plot
diagrams (see Fig. \ref{Figfr}). Evidently, the existence of root for metric
function and its number is a function of the magnetic charge. By suitable
choices of this quantity, it is possible to have naked singularity, extreme
black holes (one root) and two roots with larger root being event horizon. $%
q_{M}$ is linearly related to magnetic charge. Its value determines the
power of magnetic field. Considering this fact, Fig. \ref{Figfr} confirms a
very important fact: the super magnetized solutions suffers the absence of
horizon. In other words, the super magnetized solutions are naked
singularity.

Our final study in this section is regarding the asymptotic behavior of the
solutions. To investigate this, it is sufficient to study the behavior of
curvature scalar for large $r$. It is a matter of calculation to show that
the obtained Kretschmann scalar will have following behavior

\begin{equation}
\lim_{r\longrightarrow \infty }R_{\alpha \beta \gamma \delta }R^{\alpha
\beta \gamma \delta }\propto a_{4}\Lambda +a_{5}\Lambda ^{2}+O\left( \frac{1%
}{r^{3}}\right) ,
\end{equation}

in which $a_{4}$ and $a_{5}$ are dimension dependent coefficients.
Evidently, the dominant term in this limit is $\Lambda $ term which
indicates that the asymptotic have AdS/dS behavior depending on the sign of
cosmological constant. In the absence of the cosmological constant, the
dominant term for asymptotic behavior will be geometrical term in following
form

\begin{equation}
\lim_{\substack{ r\longrightarrow \infty  \\ \Lambda =0}}R_{\alpha \beta
\gamma \delta }R^{\alpha \beta \gamma \delta }=-\frac{a_{6}}{r^{d-1}}m+\frac{%
a_{7}}{r^{2d-4}}\left( q_{E}^{2}+q_{M}^{2}\right) +O\left( \frac{1}{r^{6}}%
\right) ,
\end{equation}%
which shows that the effects of matter field (electric and magnetic fields)
on asymptotic behavior is of secondary importance in the absence of the
cosmological constant ( $a_{6}$ and $a_{7}$ are dimension dependent
coefficients).

In conclusion, we established that these solutions enjoy the presence of a
singularity which is located at the origin and depending on choices of
different parameters, this singularity could be covered by one or two
horizons. The role of the magnetic charge on singular behavior and
properties of the horizon(s) was pointed out. In addition, the asymptotic
behavior was investigated and it was shown that in full form, it is AdS/dS
depending on negativity/positivity of the cosmological constant. In the next
section, we derive conserved quantities and study the effects of magnetic
term and higher dimensions on the thermodynamical behavior of the system.

\section{Thermodynamic properties}

In the previous section, we established the fact that our solutions could be
interpreted as black holes. Having black hole solutions, it is possible to
calculate thermodynamic quantities and study their thermodynamical behavior.
Here, our main focus is on obtaining thermodynamical properties and
employing the first law of thermodynamics to study various properties of
these black holes.

The entropy of Einsteinian black holes could be extracted by using the area
law, which leads to

\begin{equation}
S=\frac{\pi r_{+}^{d-2}}{4},  \label{entropy}
\end{equation}%
in which $r_{+}$ is the outer horizon (the largest positive real root of
metric function). The total electric charge could be obtained through the
use of the Gauss law which leads to

\begin{equation}
Q_{E}=\frac{q_{E}}{4\pi }.  \label{electric charge}
\end{equation}

The same method could be employed to calculate total magnetic charge which is

\begin{equation}
Q_{M}=\frac{q_{M}}{4\pi }.  \label{magnetic charge}
\end{equation}

It is interesting to note that although the electric and magnetic parts of $%
F_{\mu \nu }$ are completely different, their conserved charges are of the
same nature which are arisen from Gauss's law. Besides, the total mass of
these black holes could be calculated by using the ADM
(Arnowitt-Deser-Misner) method which leads to

\begin{equation}
M=\frac{\left( d-2\right) }{16}m.  \label{mass}
\end{equation}

By evaluating the metric function on the horizon, one can also find the
geometrical mass. Using obtained entropy (\ref{entropy}), electric (\ref%
{electric charge}) and magnetic charges (\ref{magnetic charge}), one can
obtain the following Smarr like formula for these black holes

\begin{equation}
M\left( S,Q_{E},Q_{m}\right) =\frac{\left( d-2\right) }{16}\left( \frac{4S}{%
\pi }\right) ^{\frac{d-3}{d-2}}k-\frac{1}{8\left( d-1\right) }\Lambda \left(
\frac{4S}{\pi }\right) ^{\frac{d-1}{d-2}}+\frac{2\pi ^{2}\left(
Q_{E}^{2}+Q_{M}^{2}\right) }{\left( d-3\right) }\left( \frac{4S}{\pi }%
\right) ^{-\frac{d-3}{d-2}}.  \label{smarr}
\end{equation}

Using the obtained mass together with the first law of black hole
thermodynamics

\begin{equation}
dM=TdS+\Phi _{E}dQ_{E}+\Phi _{M}dQ_{M},  \label{first law}
\end{equation}%
one can extract the electric and magnetic potentials respectively in the
following forms

\begin{eqnarray}
\Phi _{E} &=&\frac{\pi q_{E}}{\left( d-3\right) r_{+}^{d-3}},
\label{electric potential} \\
&&  \notag \\
\Phi _{M} &=&\frac{\pi q_{M}}{\left( d-3\right) r_{+}^{d-3}}.
\label{magnetic potential}
\end{eqnarray}

Recently, there has been a renewed proposal for the cosmological constant;
which is taken not as a fixed parameter, instead, it is regarded as a
thermodynamic variable known as dynamical pressure \cite{Kubiznak}. The
relation between these two quantities is given by

\begin{equation}
P=-\frac{\Lambda }{8\pi },
\end{equation}%
which by replacing it in Smarr like formula, one can obtain

\begin{equation}
M\left( S,Q_{E},Q_{M}\right) =\frac{\left( d-2\right) }{16}\left( \frac{4S}{%
\pi }\right) ^{\frac{d-3}{d-2}}k+\frac{\pi }{\left( d-1\right) }P\left(
\frac{4S}{\pi }\right) ^{\frac{d-1}{d-2}}+\frac{2\pi ^{2}\left(
Q_{E}^{2}+Q_{M}^{2}\right) }{\left( d-3\right) }\left( \frac{4S}{\pi }%
\right) ^{-\frac{d-3}{d-2}}.  \label{smarr extended}
\end{equation}

This consideration modifies the role of the mass from internal energy to
enthalpy and the first law of black hole thermodynamics (\ref{first law})
will be modified into

\begin{equation}
dH=TdS+\Phi _{E}dQ_{E}+\Phi _{M}dQ_{M}+VdP.  \label{first law extended}
\end{equation}

Using this relation, it is a matter of calculation to extract the
corresponding volume

\begin{equation}
V=\left( \frac{\partial H\left( S,Q_{E},Q_{M},P\right) }{\partial P}\right)
_{S,Q_{M},Q_{E}}=\left( \frac{\partial M\left( S,Q_{E},Q_{M},P\right) }{%
\partial P}\right) _{S,Q_{M},Q_{E}}=\frac{\pi r_{+}^{d-1}}{\left( d-1\right)
},  \label{volume}
\end{equation}%
which is geometrically expected. The temperature of black hole is generally
obtained through the concept of surface gravity which is given by

\begin{equation}
\kappa =\sqrt{-\frac{1}{2}\left( \nabla _{\mu }\chi _{\nu }\right) \left(
\nabla ^{\mu }\chi ^{\nu }\right) },
\end{equation}%
where $\chi ^{\nu }$ is a Killing vector. Considering the fact that our
solutions are static, the Killing vector will be $\chi =\partial _{t}$, and
therefore, temperature is calculated as

\begin{equation}
T=\frac{\kappa }{2\pi }=\frac{1}{4\pi }\frac{df\left( r\right) }{dr},
\label{temp}
\end{equation}

By replacing the cosmological constant with its correspondence pressure in
temperature (\ref{temp}), it is possible to extract an equation of state.
Using the equation of state and checking the existence of its inflection
point

\begin{equation}
\left( \frac{\partial P}{\partial r_{+}}\right) _{T}=\left( \frac{\partial
^{2}P}{\partial r_{+}^{2}}\right) _{T}=0,  \label{infel}
\end{equation}%
one is able to extract the critical points. Furthermore, it is possible to
obtain the free energy by using the following relation

\begin{equation}
F=H-TS-\Phi _{E}Q_{E}.  \label{free}
\end{equation}

The last thermodynamic quantity of interest is the heat capacity which could
be used to determine thermal stability and possible phase transition of the
black holes. This quantity is given by

\begin{equation}
C=T\left( \frac{\partial S}{\partial T}\right) _{Q_{M},Q_{E},P}.
\label{heat}
\end{equation}

In next sections, we will investigate thermodynamical behavior of the
solutions for the following ensembles; i) general case, ii) constant
electric field, iii) constant magnetic field, and iv) constant electric and
magnetic fields.

Unfortunately, it was not possible to extract all the properties for
mentioned cases in arbitrary dimension, analytically. Therefore, we consider
$5$-dimensional black holes (as a prototype higher dimensional solutions) as
a case study to understand the effects of higher dimensions on properties of
solutions.

\section{General case}

Here, we consider unspecified electric and magnetic fields and study the
general behavior. Using obtained mass (\ref{smarr extended}), it is a matter
of calculation to show that mass/enthalpy for this case is given by

\begin{equation}
M=H=\frac{\left( d-2\right) r_{+}^{d-3}}{16}k+\frac{\pi r_{+}^{d-1}}{%
16\left( d-1\right) }P+\frac{r_{+}^{3-d}}{8\left( d-3\right) }\left(
q_{E}^{2}+q_{M}^{2}\right) .  \label{mass case 1}
\end{equation}

As it was pointed out, it is not possible to obtain roots of the
mass/enthalpy for arbitrary $d$, analytically. Therefore, we consider $5$%
-dimensional solution as a case study. So, in 5-dimensions, the root is
given by

\begin{equation}
r_{+}(M(d=5)=0)=\frac{\sqrt{A_{1}^{2/3}+k^{2}-kA_{1}^{1/3}}}{2\sqrt{\pi P}%
A_{1}^{1/6}},
\end{equation}%
in which $A_{1}=4\pi P\sqrt{q_{E}^{2}+q_{M}^{2}}\sqrt{4\pi ^{2}P^{2}\left(
q_{E}^{2}+q_{M}^{2}\right) +k^{3}}-8\pi ^{2}P^{2}\left(
q_{E}^{2}+q_{M}^{2}\right) -k^{3}$.

In addition, the high energy limit and asymptotical behavior of the mass are
given by

\begin{eqnarray}
\lim_{r_{+}\rightarrow 0}M &=&\frac{1}{16r_{+}^{2}}\left(
q_{E}^{2}+q_{M}^{2}\right) +\frac{3}{16}r_{+}^{2}k+O\left( r_{+}^{4}\right) ,
\\
&&  \notag \\
\lim_{r_{+}\rightarrow \infty }M &=&\frac{\pi }{4}r_{+}^{4}P+\frac{3}{16}%
r_{+}^{2}k+O\left( \frac{1}{r_{+}^{2}}\right) .
\end{eqnarray}

The high energy limit of the mass (enthalpy) for this case is governed by
magnetic and electric charge terms. This indicates that in the absence of
the electric field, the dominant term in the high energy limit of the mass
is magnetic charge of the solutions. The effects of electric and magnetic
charges are of the same order. On the other hand, the asymptotic behavior is
governed by the pressure term which is essentially the cosmological constant
term. In general, one can state that for small black holes, the effects of
matter field, hence the magnetic and electric charges become dominant over
other quantities contributing to the mass, whereas for large black holes,
the significant effect comes from the pressure term. This difference in the
limiting behavior could be employed to determine the size of black hole,
although there are other matters that should be considered before making a
final statement. But, the important subject is that here, the contribution
of the magnetic charge becomes significant as we study the high energy
limit. Considering that for both limits, the dominant terms are positive
valued with power of the horizon radius presented for different terms, one
expects a minimum for the mass. Plotting mass versus $r_{+}$ confirms this
statement (see left panel of Fig. \ref{FigEM1}). The minimum takes place
where the topological term, $k$, becomes dominant which is the case for
medium black holes. This indicates that the smallest mass (enthalpy)
available for these black holes belongs to medium black holes.

Using Eq. (\ref{temp}) or $\left( \frac{\partial H}{\partial S}\right)
_{q_{E},q_{M}}$, it is possible to extract the temperature as

\begin{equation}
T=\frac{1}{2\pi }\left[ \frac{\left( d-3\right) }{2r_{+}}k+\frac{8\pi r_{+}}{%
\left( d-2\right) }P-\frac{q_{E}^{2}+q_{M}^{2}}{\left( d-2\right)
r_{+}^{2d-5}}\right] .  \label{temperature case1}
\end{equation}

The root of this quantity could not be calculated for $d$-dimensional case,
whereas for the $5$-dimensional case it is obtained as

\begin{equation}
r_{+}(T(d=5)=0)=\frac{\sqrt{A_{2}^{2/3}+k^{2}-kA_{2}^{1/3}}}{\sqrt{8\pi P}%
A_{2}^{1/6}},
\end{equation}%
in which $A_{2}=8\pi P\sqrt{q_{E}^{2}+q_{M}^{2}}\sqrt{16\pi ^{2}P^{2}\left(
q_{E}^{2}+q_{M}^{2}\right) +k^{3}}+32\pi ^{2}P^{2}\left(
q_{E}^{2}+q_{M}^{2}\right) -k^{3}$. The high energy limit and asymptotic
behavior of the temperature are given by

\begin{eqnarray}
\lim_{r_{+}\rightarrow 0}T &=&-\frac{1}{6\pi r_{+}^{5}}\left(
q_{E}^{2}+q_{M}^{2}\right) +\frac{1}{2\pi r_{+}}k+O\left( r_{+}\right) , \\
&&  \notag \\
\lim_{r_{+}\rightarrow \infty }T &=&\frac{4}{3}r_{+}P+\frac{1}{2\pi r_{+}}%
k+O\left( \frac{1}{r_{+}^{5}}\right) .
\end{eqnarray}

\begin{figure}[tbp]
$%
\begin{array}{ccc}
\epsfxsize=5.75cm \epsffile{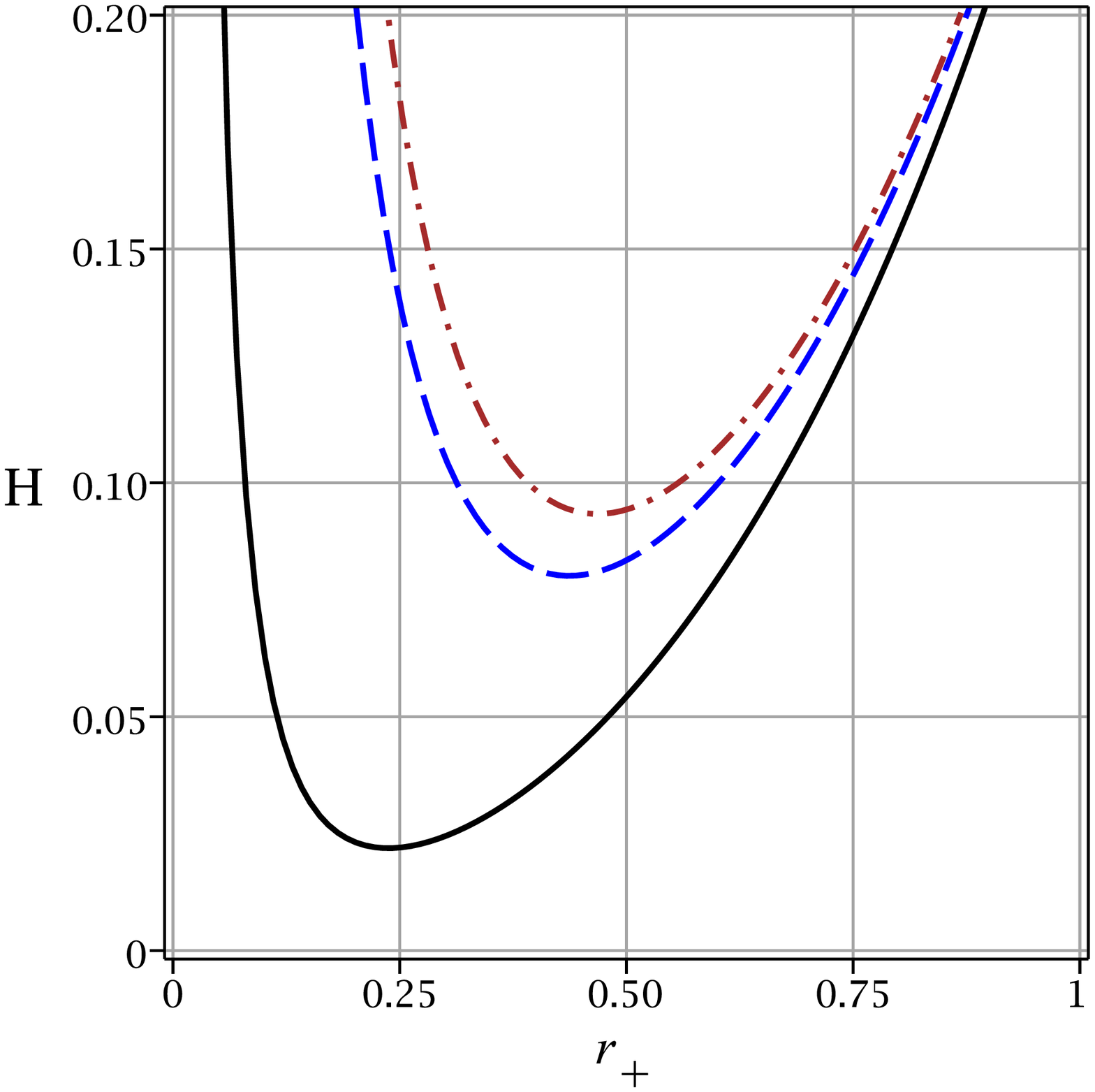} & \epsfxsize=5.75cm \epsffile{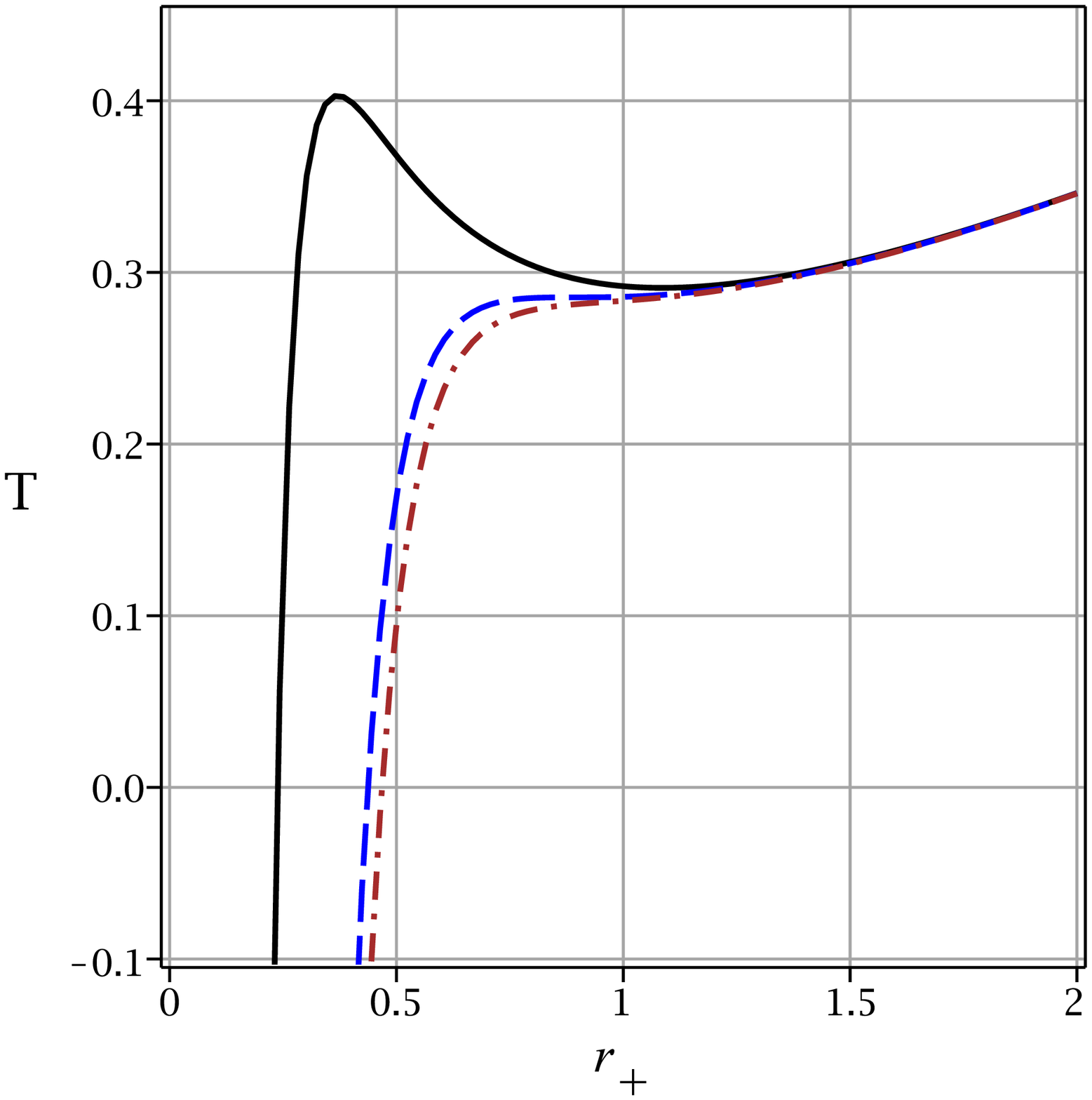}
& \epsfxsize=5.75cm \epsffile{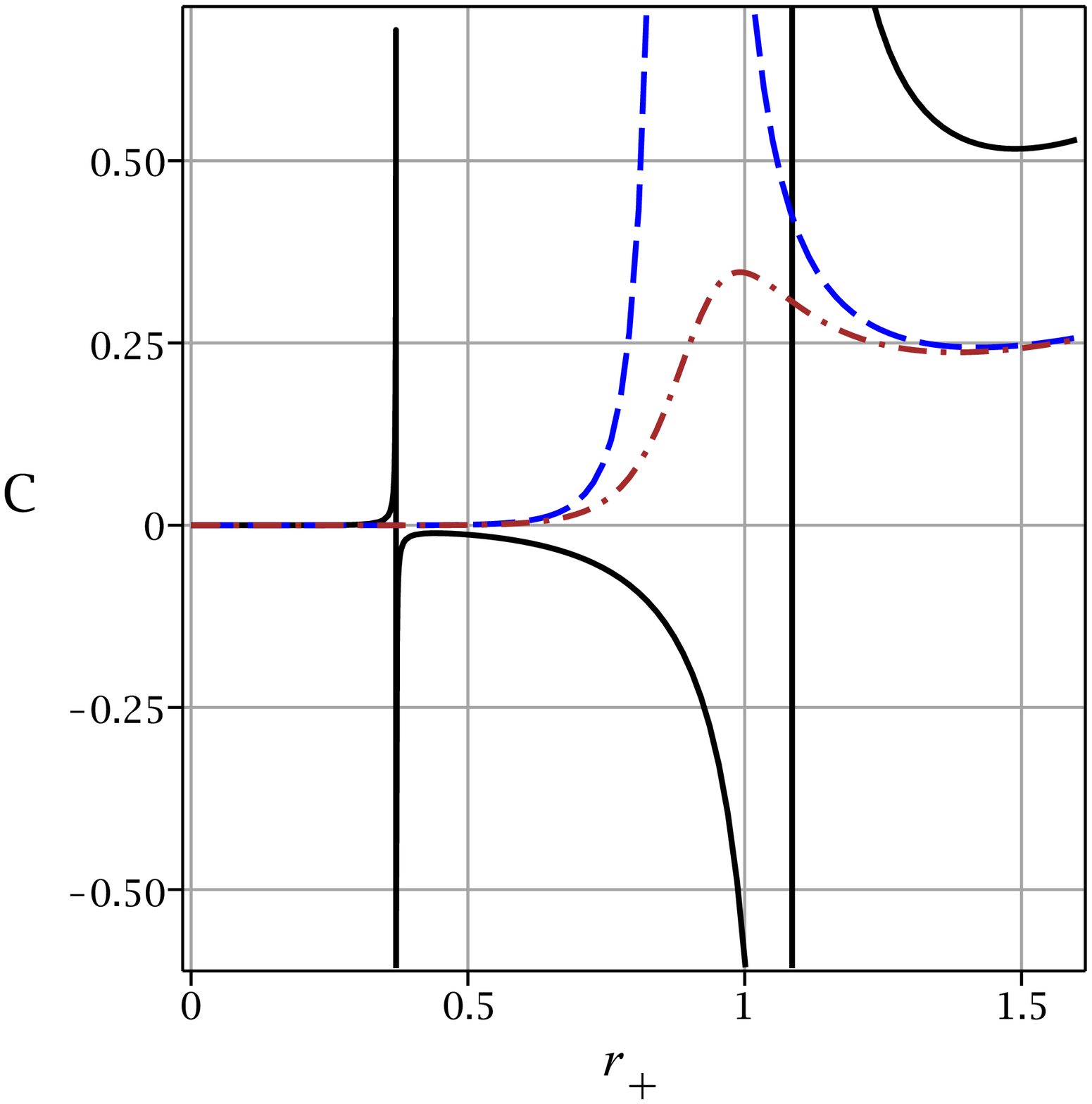}%
\end{array}
$%
\caption{$H$ (left panel), $T$ (middle panel) and $C$ (right panel) versus $%
r_{+}$ for $d=5$, $k=1$, $P=0.1$ and $q_{E}=0.1$; $q_{M}=0$ (continuous
line), $q_{M}=0.34155$ (dashed line) and $q_{M}=0.4$ (dashed-dotted line).}
\label{FigG1}
\end{figure}

Here too, similar to enthalpy, the dominant term in high energy limit
includes electric and magnetic charges but contrary to enthalpy case, its
sign is negative. On the other hand, the asymptotic behavior of the
temperature is governed by the pressure term whereas for medium black holes,
the topological factor determines the behavior of temperature. Since the
high energy limit has negative sign and asymptotic behavior is positive, it
is expected that there is at least one real valued positive root for
temperature which could be observed in plotted diagrams (see middle panel of
Fig. \ref{FigG1}). Studying the diagrams for temperature confirms that
depending on the choices of magnetic charge, temperature: i) could be an
increasing function of horizon radius with one root. ii) could have one
extremum and one root. iii) could have one root with one minimum and one
maximum which are located after the root. Existence of the extremum for
temperature confirms the existence of divergency for the heat capacity. In
other words, extrema are where the heat capacity acquires divergencies, and
therefore, they are phase transition points. Here, we see that modification
in magnetic charge results in existence/absence of extremum for the
temperature. In addition, since before root, the temperature is negative and
solutions are not physical, one can conclude that modification in magnetic
charge changes the valid range for existence of physical black holes. The
root of temperature is an increasing function of the magnetic charge.

It is a matter of calculation to obtain the equation of state for this case
as

\begin{equation}
P=\frac{1}{2\pi }\left[ \frac{\pi \left( d-2\right) }{2r_{+}}T-\frac{\left(
d-2\right) \left( d-3\right) }{8r_{+}^{2}}k+\frac{q_{E}^{2}+q_{M}^{2}}{%
4r_{+}^{2d-4}}\right] .  \label{pressure case 1}
\end{equation}

Using the concept of inflection point (\ref{infel}), one can obtain the
following equation for calculating the critical horizon radius

\begin{equation}
\frac{\left( q_{E}^{2}+q_{M}^{2}\right) \left( 4d-10\right) }{r_{+}^{2d-6}}%
-\left( d-3\right) k=0,
\end{equation}%
in which the critical horizon radius, temperature and pressure are,
respectively, obtained as

\begin{equation}
r_{c}=\left[ \frac{\left( q_{E}^{2}+q_{M}^{2}\right) \left( 4d-10\right) }{%
\left( d-3\right) k}\right] ^{\frac{1}{2d-6}},
\end{equation}

\begin{equation}
T_{c}=\frac{1}{\pi }\left( \frac{k}{2d-5}\right) ^{\frac{2d-5}{2d-6}}\left(
d-3\right) ^{\frac{4d-11}{2d-6}}\left( 2q_{E}^{2}+2q_{M}^{2}\right) ^{-\frac{%
1}{2d-6}},
\end{equation}

\begin{eqnarray}
P_{c} &=&\frac{k^{\frac{d-2}{d-3}}}{16\pi \left( q_{E}^{2}+q_{M}^{2}\right)
^{1/\left( d-3\right) }}\left\{ 2^{\frac{2d-7}{d-3}}\left( 2d-5\right) ^{%
\frac{2-d}{d-3}}\left[ \left( d-2\right) \left( d-3\right) ^{\frac{2d-5}{d-3}%
}+\frac{\left( d-3\right) ^{\frac{d-2}{d-3}}}{4}\right] \right.  \notag \\
&&\left. -\left( \frac{\left( d-3\right) ^{d-2}}{2d-5}\right) ^{1/\left(
d-3\right) }\left[ \frac{d}{2^{1/\left( d-3\right) }}-2^{\frac{d-4}{d-3}}%
\right] \right\} .
\end{eqnarray}

\begin{figure}[tbp]
$%
\begin{array}{c}
\epsfxsize=7cm \epsffile{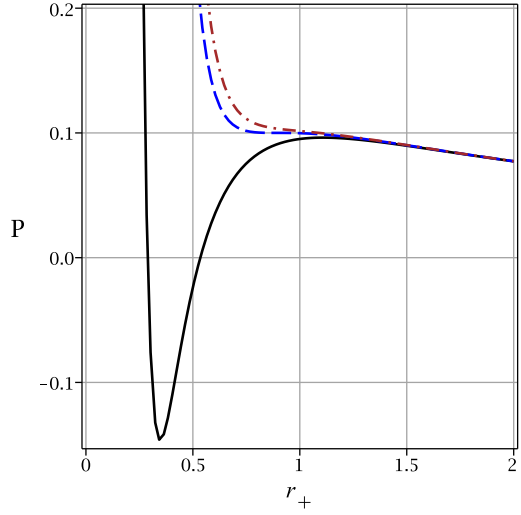}%
\end{array}
$%
\caption{$P$ versus $r_{+}$ for $d=5$, $k=1$, $T=0.2855$ and $q_{E}=0.1$; $%
q_{M}=0$ (continuous line), $q_{M}=0.34155$ (dashed line) and $q_{M}=0.4$
(dashed-dotted line).}
\label{FigG2}
\end{figure}

Evidently, the critical horizon radius is an increasing function of the
magnetic charge while critical temperature and pressure are decreasing
function of it. This indicates that for magnetized black holes, phase
transition takes place in larger horizon radius for smaller pressure and
temperature. Therefore, one can conclude that for super magnetized black
holes (large magnetic charge), van der Waals like phase transition takes
place in very small values of temperature and pressure but for large volume.
This enables one to understand the effects of magnetization on van der Waals
like behavior of the black holes. In order to confirm the existence of van
der Waals like behavior, we have plotted the following diagrams for the
pressure (see Fig. \ref{FigG2}). Evidently, by variation of the magnetic
charge, the pressure can acquire i) one extremum which indicates existence
of critical behavior taking place at one point. ii) two extrema which show
the existence of phase transition over range (rather than a single point)
iii) without any extremum which is interpreted as the absence of critical
behavior.

Using obtained thermodynamical quantities, it is possible to calculate the
free energy as

\begin{equation}
F=\frac{r_{+}^{d-3}}{16}k-\frac{\pi r_{+}^{d-1}}{\left( d-1\right) \left(
d-2\right) }P+\frac{\left( 2d-5\right) r_{+}^{d-3}}{8\left( d-2\right)
\left( d-3\right) }q_{M}^{2}+\frac{2\left( 2d-5\right) \pi -\left(
d-2\right) }{16\pi \left( d-2\right) \left( d-3\right) }q_{E}^{2}.
\label{free case 1}
\end{equation}

Considering that $q_{E}$ term is horizon radius independent, for high energy
limit, this term becomes dominant, whereas, for asymptotic behavior, the
dominant term is pressure term. Finally, the heat capacity of this case is
obtained as

\begin{equation}
C=\frac{\left( d-2\right) \left( d-3\right) \pi r_{+}^{3d-2}k+16\pi
^{2}r_{+}^{3d}P-2\pi r_{+}^{d+4}\left( q_{E}^{2}+q_{M}^{2}\right) }{64\pi
r_{+}^{2d+2}P-4\left( d-2\right) \left( d-3\right) r_{+}^{2d}k+8\left(
2d-5\right) \left( q_{E}^{2}+q_{M}^{2}\right) r_{+}^{6}}\left( d-2\right) .
\label{heat capacity case 1}
\end{equation}

Divergencies of the heat capacity could not be extracted analytically for
arbitrary dimension, while for the $5$-dimensional case, divergency occurs at

\begin{equation}
r_{+}(C(d=5)\longrightarrow \infty )=\frac{\sqrt{%
A_{3}^{2/3}+k^{2}-kA_{3}^{1/3}}}{\sqrt{8\pi P}A_{3}^{1/6}}.
\end{equation}%
where $A_{3}=8\sqrt{5}\pi P\sqrt{q_{E}^{2}+q_{M}^{2}}\sqrt{80\pi
^{2}P^{2}\left( q_{E}^{2}+q_{M}^{2}\right) -k^{3}}-160\pi ^{2}P^{2}\left(
q_{E}^{2}+q_{M}^{2}\right) +k^{3}$.

In addition, based on series expansion, we find that high energy limit and
asymptotic behavior of the heat capacity are given by

\begin{eqnarray}
\lim_{r_{+}\rightarrow 0}C &=&-\frac{3\pi }{20}r_{+}^{3}+\frac{9}{25}\frac{%
k\pi }{q_{E}^{2}+q_{M}^{2}}r_{+}^{7}+O\left( r_{+}^{9}\right) , \\
&&  \notag \\
\lim_{r_{+}\rightarrow \infty }C &=&\frac{3\pi }{4}r_{+}^{3}+\frac{9}{16}%
\frac{k}{P}r_{+}+O\left( \frac{1}{r_{+}^{3}}\right) .
\end{eqnarray}

The effect of the magnetic charge on the obtained divergency for the heat
capacity is evident. This shows that phase transition points, root and
stability conditions of the black holes are modified due to the magnetic
charge. If certain conditions are satisfied, the heat capacity could acquire
two divergencies (see right panel of Fig. \ref{FigG1}). Between the
divergencies, solutions suffer from thermal instability, since the heat
capacity is negative valued. But comparing this case with plotted diagrams
for the pressure, one can see that extrema of pressure coincide with
divergencies of the heat capacity. Therefore, the region between
divergencies is where physical black holes are absent. As a result, one can
draw the following conclusions regarding thermal stability of the solutions:
i) in the absence of divergencies, stable black holes exist after root of
the heat capacity which is the same as root of the temperature. ii) in the
case of one divergency, there is a phase transition between medium stable
black holes and larger stable ones taking place at the divergency. Before
root of the heat capacity (temperature), solutions are unstable and
non-physical (due to negative temperature). iii) in case of two divergencies
for the heat capacity, the phase transition takes place over a region
between small stable black holes and large stable ones. Once more, we
emphasize that between two divergencies, due to thermodynamical concepts, no
physical black holes exist.

Interestingly, the high energy limit of heat capacity and asymptotic
behavior of it depends only on horizon radius with some factors. In other
words, the high energy limit and asymptotic behavior of the heat capacity
are depending neither on electric and magnetic charges nor on pressure. It
seems that for the heat capacity, these two limits are only affected by the
gravitational part of the action, since there is no trace of matter field
and cosmological constant generalization present in them. It is worthwhile
to mention that the second dominant term in high energy limit of the heat
capacity includes electric and magnetic charges which highlights the
contribution of magnetic charge.


\section{Constant electric field}

In this section, we work in an ensemble where the temporal component of the
electromagnetic tensor, $A_{t}$, is constant everywhere. Therefore, we can
replace the electric charge, $q_{E},$ by its value at the horizon with
following relation

\begin{equation}
q_{E}=\frac{\left( d-3\right) r_{+}^{d-3}\Phi _{E}}{\pi }.
\end{equation}

So, the mass/enthalpy of this case is given by

\begin{equation}
M=\frac{\left( d-2\right) r_{+}^{d-3}}{16}k+\frac{\pi r_{+}^{d-1}}{d-1}P+%
\frac{r_{+}^{3-d}}{8\left( d-3\right) }q_{M}^{2}+\frac{\left( d-3\right)
r_{+}^{d-3}}{8\pi ^{2}}\Phi _{E}^{2},  \label{mass case 2}
\end{equation}%
in which, its roots could not be obtained analytically, except for the $5$%
-dimensional case which are

\begin{equation}
r_{+}(M(d=5)=0)=\frac{1}{6}\sqrt{\frac{\left( \frac{27\left( \pi ^{2}k+\frac{%
4\Phi _{E}^{2}}{3}\right) ^{2}}{A_{4}^{1/3}}+3A_{4}^{1/3}-12\Phi
_{E}^{2}-9\pi ^{2}k\right) }{\pi ^{3}P}},
\end{equation}%
where $A_{4}=-A_{5}+12\sqrt{3}\pi ^{4}Pq_{M}\sqrt{A_{5}}$ and $A_{5}=216\pi
^{8}P^{2}q_{M}^{2}+27\pi ^{6}k^{3}+108\pi ^{4}k^{2}\Phi _{E}^{2}+144\pi
^{2}k\Phi _{E}^{4}+64\Phi _{E}^{6}$. Also the high energy limit and
asymptotic behavior are given by

\begin{eqnarray}
\lim_{r_{+}\rightarrow 0}M &=&\frac{1}{16r_{+}^{2}}q_{M}^{2}+\left( \frac{3}{%
16}k+\frac{1}{4\pi ^{2}}\Phi _{E}^{2}\right) r_{+}^{2}+O\left(
r_{+}^{4}\right) , \\
&&  \notag \\
\lim_{r_{+}\rightarrow \infty }M &=&\frac{\pi }{4}r_{+}^{4}P+\left( \frac{3}{%
16}k+\frac{1}{4\pi ^{2}}\Phi _{E}^{2}\right) r_{+}^{2}+O\left( \frac{1}{%
r_{+}^{2}}\right) .
\end{eqnarray}

\begin{figure}[tbp]
$%
\begin{array}{ccc}
\epsfxsize=5.75cm \epsffile{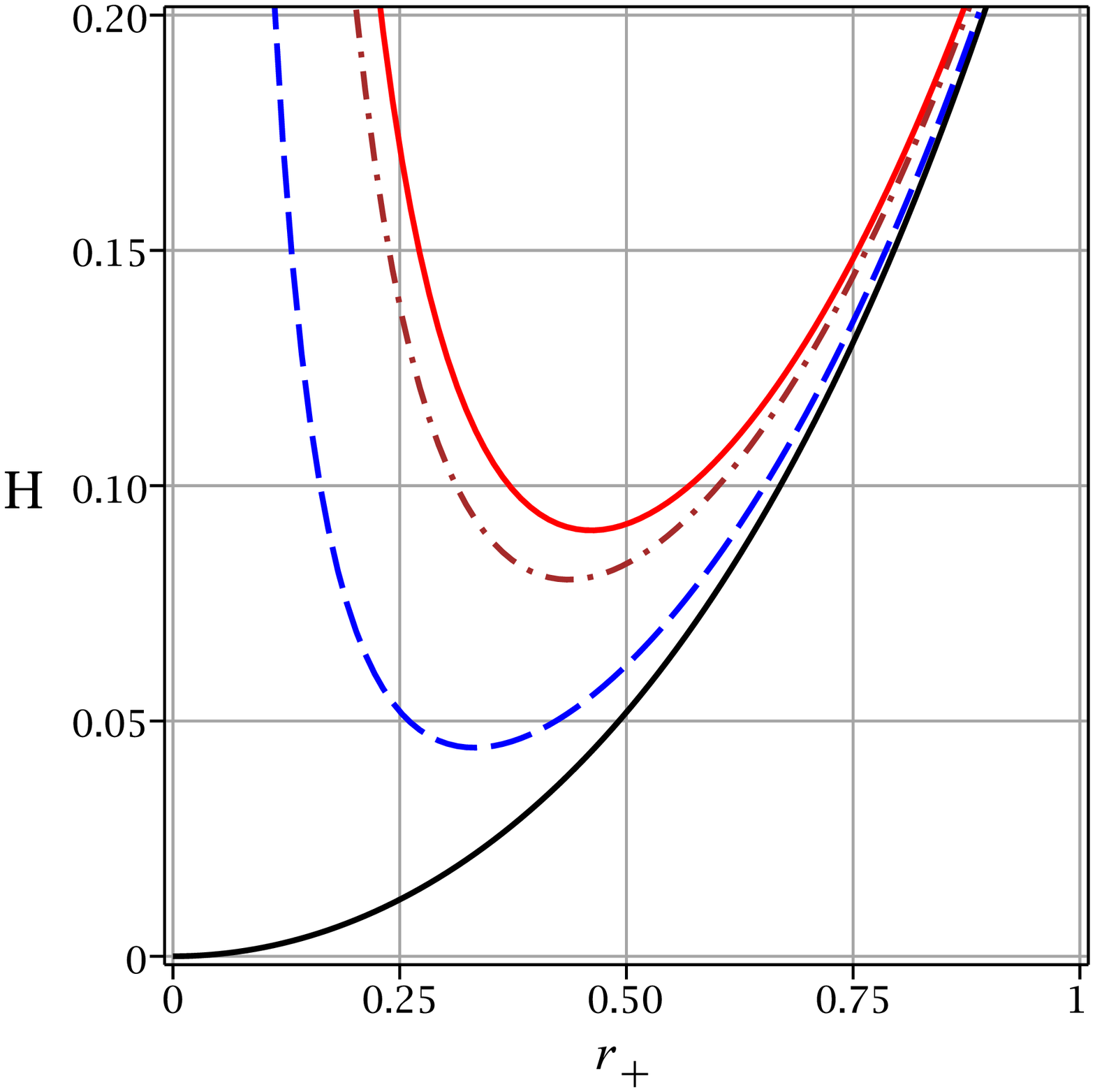} & \epsfxsize=5.75cm \epsffile{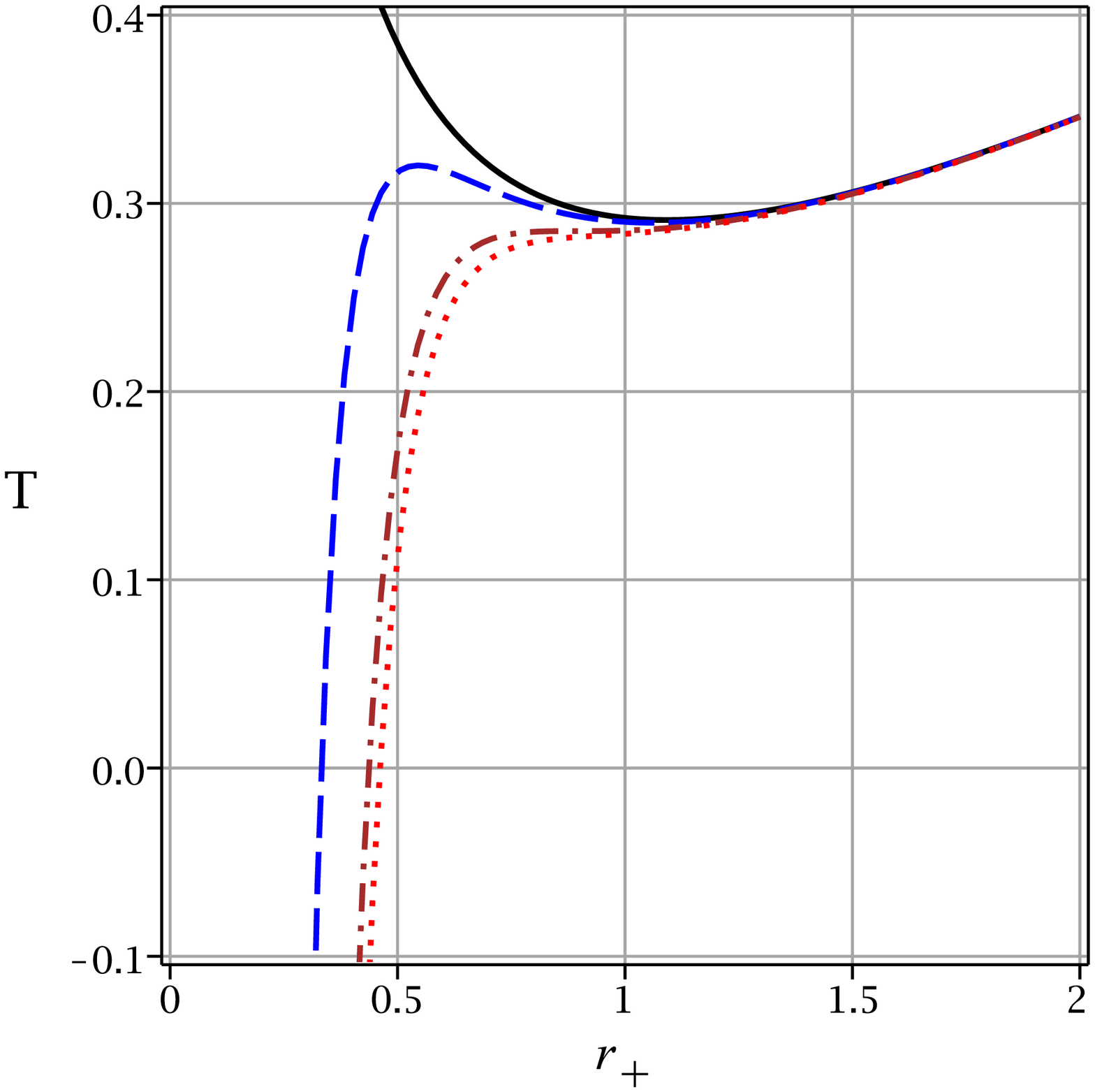}
& \epsfxsize=5.75cm \epsffile{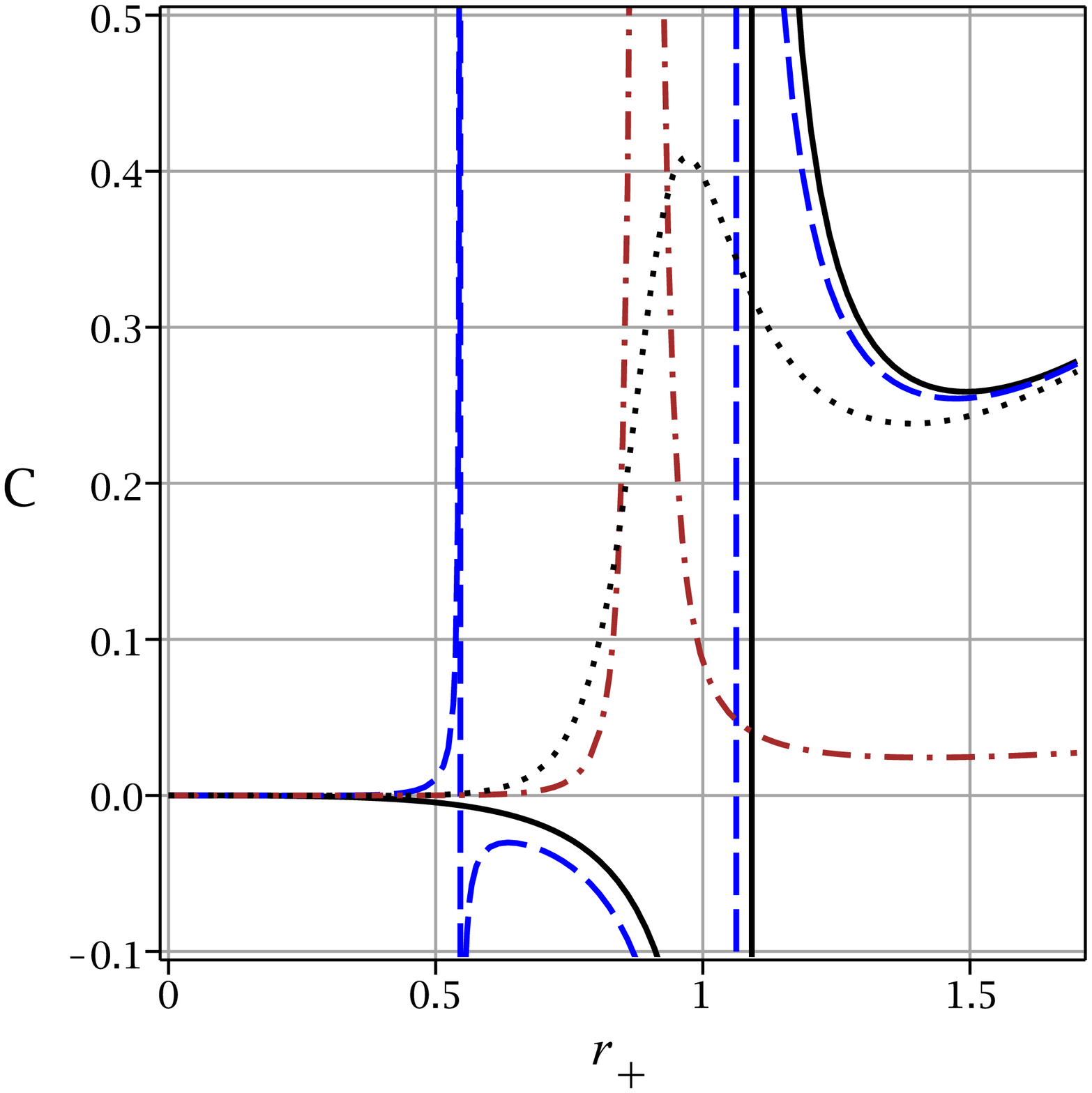}%
\end{array}
$%
\caption{$H$ (left panel), $T$ (middle panel) and $C$ (right panel) versus $%
r_{+}$ for $d=5$, $k=1$, $P=0.1$ and $\Phi _{E}=0.1$; $q _{M}=0$ (continuous
line), $q _{M}=0.2$ (dashed line), $q _{M}=0.3553$ (dashed-dotted line) and $%
q _{M}=0.4$ (dotted line).}
\label{FigE1}
\end{figure}

The high energy limit of mass/enthalpy of this case is governed by the
magnetic charge. This means that for small black holes the magnetization has
a significant role on the behavior of enthalpy. That being said, one can see
that due to the presence of magnetic charge (none constant magnetic field),
the enthalpy for vanishing the horizon radius diverges. Comparing this case
with previous one, it can be stated that the presence of magnetic charge
with constant electric field leads to the high energy limit of enthalpy
becomes modified at a significant level. The enthalpy is an increasing
function of the magnetic charge. Taking a closer look at the high energy
limit and asymptotic behavior, one can see that for medium range of horizon
radius dominant terms of enthalpy are both electric potential and
topological term. Interestingly, by choosing

\begin{equation}
\Phi _{E}=\pi \sqrt{-\frac{3k}{4}},
\end{equation}%
for $k=-1$, it is possible to cancel out all the effects of electric field
on enthalpy. It is clear that such a situation could only occur for black
holes with hyperbolic horizon. This specific behavior is rooted in the
assumption of electric field being constant. In order to have a better
picture regarding thermodynamic behavior of the enthalpy for this case, we
have plotted qa diagram (see left panel of Fig. \ref{FigE1}). Evidently, for
constant electric field and in the absence of magnetic charge, enthalpy is
an increasing function of horizon radius. The situation is modified in the
presence of magnetic charge. In this case, the enthalpy acquires a minimum.
Depending on choices of different parameters (specially topological factor),
the enthalpy can have no root, one root, or two roots. It is notable that
existence of two roots and presence of a minimum originates from
contributions of the magnetic charge.

The temperature for this case is obtained as

\begin{equation}
T=\frac{1}{2\pi }\left[ \frac{\left( d-3\right) }{2r_{+}}k+\frac{8\pi r_{+}}{%
\left( d-2\right) }P-\frac{q_{M}^{2}}{\left( d-2\right) r_{+}^{2d-5}}-\frac{%
\left( d-3\right) ^{2}}{\pi ^{2}\left( d-2\right) r_{+}}\Phi _{E}^{2}\right]
.
\end{equation}

The root of the temperature could not be obtained analytically for general $%
d $-dimensional case, but in $5$- dimensions, it is given by

\begin{equation}
r(T(d=5)=0)=\frac{1}{12}\sqrt{\frac{\left( \frac{54\left( \pi ^{2}k-\frac{%
4\Phi _{E}^{2}}{3}\right) ^{2}}{A_{6}^{1/3}}+6A_{6}^{1/3}+24\Phi
_{E}^{2}-18\pi ^{2}k\right) }{\pi ^{3}P}},
\end{equation}%
where $A_{6}=A_{7}+24\sqrt{3}\pi ^{4}Pq_{M}\sqrt{A_{8}}$, in which $A_{7}$
and $A_{8}$ are

\begin{eqnarray}
A_{7} &=&864\pi ^{8}P^{2}q_{M}^{2}-27\pi ^{6}k^{3}+108\pi ^{4}k^{2}\Phi
_{E}^{2}-144\pi ^{2}k\Phi _{E}^{4}+64\Phi _{E}^{6}, \\
&&  \notag \\
A_{8} &=&432\pi ^{8}P^{2}q_{M}^{2}-27\pi ^{6}k^{3}+108\pi ^{4}k^{2}\Phi
_{E}^{2}-144\pi ^{2}k\Phi _{E}^{4}+64\Phi _{E}^{6}.
\end{eqnarray}

For high energy limit and asymptotic behavior we have

\begin{eqnarray}
\lim_{r_{+}\rightarrow 0}T &=&-\frac{1}{6\pi r_{+}^{5}}q_{M}^{2}+\left(
\frac{1}{2}k-\frac{2}{3\pi ^{3}}\Phi _{E}^{2}\right) \frac{1}{r_{+}}+O\left(
r_{+}\right) , \\
&&  \notag \\
\lim_{r_{+}\rightarrow \infty }T &=&\frac{4}{3}r_{+}P+\left( \frac{1}{2\pi }%
k-\frac{2}{3\pi ^{3}}\Phi _{E}^{2}\right) \frac{1}{r_{+}}+O\left( \frac{1}{%
r_{+}^{5}}\right) .
\end{eqnarray}

The dominant term in the high energy limit is the magnetic term with
negative sign. For medium black holes, topological and electric potential
terms are governing the behavior of temperature. Finally, the asymptotic
behavior is governed by pressure, hence cosmological term. It is worthwhile
to mention that for $\Phi =\pi \sqrt{\frac{3k}{4}}$, the effects of
topological and electric potential cancel each other. This could only take
place for spherical black holes ($k=1$). Since the high energy limit of the
temperature is negative valued and asymptotic behavior is positive, one can
confirm that there exists at least one root for the temperature. To show
this, we have plotted it in middle panel of Fig. \ref{FigE1}. In the absence
of magnetic charge, temperature has a minimum. By taking a non-zero value
for the magnetic charge, one can find the following behavior for
temperature: There exists a critical magnetic charge, $q_{M-critical}$ at
which the temperature acquires one extremum. For magnetic charges less than
this critical magnetic charge, $q_{M}<q_{M-critical}$, the temperature has
two exterma: one minimum and one maximum. On the other hand, for $%
q_{M}>q_{M-critical}$, no extremum is available for the temperature.
Remembering that extrema are where the heat capacity diverges, one can state
that the presence of magnetic charge enables the possibility of existence of
van der Waals like behavior for these black holes, provided by mentioned
cases. In the absence of magnetic charge, temperature indeed has a minimum,
but the type of phase transition for these two cases are different: while
one is van der Waals like phase transition the other one (absence of
magnetic charge) does not enjoy this type of phase transition. This
highlights the importance of the contribution of the magnetic charge. It is
worthwhile to mention that there exists a root for temperature which is an
increasing function of the magnetic charge.

The equation of state for the pressure is obtained in the following form

\begin{equation}
P=\frac{1}{2\pi }\left[ \frac{\pi \left( d-2\right) }{2r_{+}}T-\frac{\left(
d-2\right) \left( d-3\right) }{8r_{+}^{2}}k+\frac{q_{M}^{2}}{4r_{+}^{2d-4}}+%
\frac{\left( d-3\right) ^{2}}{4\pi ^{2}r_{+}^{2}}\Phi _{E}^{2}\right] .
\end{equation}

Using the properties of the inflection point, one can derive the following
relation governing the critical horizon radius

\begin{equation}
\frac{\left( d-2\right) \left( 4d-10\right) }{r_{+}^{2d-6}}\pi
^{2}q_{M}^{2}-\left( d-3\right) \left[ k\left( d-2\right) \pi ^{2}-2\left(
d-3\right) \Phi _{E}^{2}\right] =0,
\end{equation}%
which leads to

\begin{equation}
r_{c}=\left[ \frac{\left( d-3\right) \left[ k\left( d-2\right) \pi
^{2}-2\left( d-3\right) \Phi _{E}^{2}\right] }{2\left( d-2\right) \left(
2d-5\right) \pi ^{2}q_{M}^{2}}\right] ^{\frac{1}{2d-6}},
\end{equation}

It is a matter of calculation to obtain the critical temperature and
pressure as

\begin{eqnarray}
T_{c} &=&\left( 2d-4\right) ^{\frac{7-2d}{2d-6}}\left[ \left( 2d-5\right)
q_{M}^{2}\right] ^{\frac{1}{2d-6}}\pi ^{\frac{10-3d}{d-3}}\left[ \left(
d-2\right) \left( d-3\right) \pi ^{2}k-2\left( d-3\right) ^{2}\Phi _{E}^{2}%
\right] ^{\frac{5-2d}{2d-6}}  \notag \\
&&  \notag \\
&&\left[ \pi ^{4}\left( d-2\right) ^{2}\left( d-3\right) ^{2}k^{2}-4\left(
d-2\right) \left( d-3\right) ^{3}\pi ^{2}\Phi _{E}^{2}k-4\pi ^{4}\left(
d-2\right) ^{2}\left( 2d-5\right) q_{M}^{4}+4\left( d-3\right) ^{4}\Phi
_{E}^{4}\right] ,
\end{eqnarray}

\begin{eqnarray}
P_{c} &=&\frac{\left( \Phi _{E}^{2}\left( 6-2d\right) +\left( d-2\right) \pi
^{2}k\right) ^{\frac{2-d}{d-3}}}{4\left( d-3\right) ^{\frac{d-2}{d-3}}}%
\left\{ q_{M}^{\frac{2d-4}{d-3}}\left( q_{E}^{2}+q_{M}^{2}\right) \left[
2\pi ^{d-1}\left( d-2\right) ^{d-2}\left( 2d-5\right) ^{d-2}\right]
^{1/\left( d-3\right) }\right.  \notag \\
&&  \notag \\
&&\left. +\left[ \left( d-2\right) \left( 2d-5\right) \right] ^{1/\left(
d-3\right) }\left[ 2^{\frac{4-d}{d-3}}\Gamma -20d^{3}\pi ^{\frac{d-1}{d-3}%
}\left( k^{2}\left( \frac{q_{M}^{2}}{2^{2d-7}}\right) ^{1/\left( d-3\right)
}+\frac{\left( 2q_{M}^{4d-10}\right) ^{1/\left( d-3\right) }}{5}\right) %
\right] \right\}  \notag \\
&&  \notag \\
&&-\frac{1}{16}\left( \frac{2\pi ^{5-d}\left( d-2\right) ^{d-2}\left(
d-3\right) ^{d-4}\left( 2d-5\right) kq_{M}^{2}}{\Phi _{E}^{2}\left(
6-2d\right) +\pi ^{2}k\left( d-2\right) }\right) ^{1/\left( d-3\right) },
\end{eqnarray}%
in which

\begin{eqnarray*}
\Gamma &=&52\left( \frac{13d^{2}-28d+20}{13}\right) \left( \pi
^{d-1}q_{M}^{4d-10}\right) ^{1/\left( d-3\right) }-4k\Phi _{E}^{2}\left(
d-2\right) \left( d-3\right) ^{3}\left( \frac{q_{M}^{2}}{\pi ^{d-5}}\right)
^{1/\left( d-3\right) } \\
&&+\left( \frac{4\left( d-3\right) ^{4}\Phi _{E}^{2}}{\pi ^{\frac{3d-11}{d-3}%
}}+\pi ^{\frac{d-1}{d-3}}k^{2}\left( d^{4}+37d^{2}-60d+36\right) \right)
q_{M}^{\frac{4d-10}{d-3}}
\end{eqnarray*}

\begin{figure}[tbp]
$%
\begin{array}{c}
\epsfxsize=7cm \epsffile{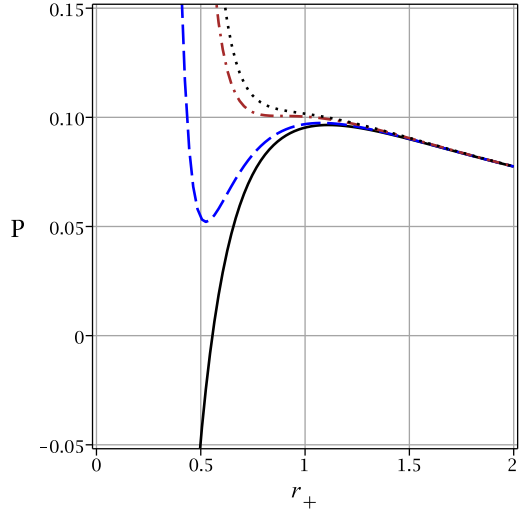}%
\end{array}
$%
\caption{$P$ versus $r_{+}$ for $d=5$, $k=1$, $T=0.286$ and $\Phi _{E}=0.1$;
$q _{M}=0$ (continuous line), $q _{M}=0.2$ (dashed line), $q _{M}=0.3553$
(dashed-dotted line) and $q _{M}=0.4$ (dotted line).}
\label{FigE2}
\end{figure}

Evidently, here, the critical horizon radius is a decreasing function of the
electric potential and magnetic charge while it is an increasing function of
the topological parameter (if taken as a continuous variable). In previous
case, the critical horizon radius was a decreasing function of the electric
charge but the presence of the electric charge was in the denominator of the
critical horizon radius. But here, the electric potential is in the
numerator of the obtained critical horizon radius with negative sign. The
positive real valued critical horizon radius only exits for the spherical
case. On the other hand, critical temperature and pressure are decreasing
functions of the magnetic charge and electric potential. Here, the obtained
critical temperature and pressure show a significant modification compared
to the previous case. This indicates that consideration of the electric
field being constant resulted in different classes of black holes which have
different thermodynamical properties, therefore critical structure. We
should point it out that in the absence of electric field, the black holes
will have critical behavior which indicates that magnetic black holes have
also van der Waals like behavior in their structure considering the set up
proposed in this paper. In order to have a better picture regarding the
effects of magnetic charge on this case, we have plotted Fig. \ref{FigE2}.
Evidently, in the absence of magnetic charge, no van der Waals like behavior
is present for the pressure vs horizon radius. In this case, pressure has a
root and maximum which appears after root. Therefore, for this case, there
is a region of negative pressure before root. After such root, there exists
a maximum which marks the existence of a phase transition which is not van
der Waals like. Later, in studying the heat capacity diagrams, we will see
what type of phase transition these diagrams represent. In the presence of
magnetic charge, the behavior of pressure diagrams is modified. There exists
a critical magnetic charge, $q_{M-critical}$ in which for $%
q_{M}<q_{M-critical}$, pressure has one minimum and one maximum. The van der
Waals like phase transition takes place over this region. For $%
q_{M}=q_{M-critical}$, the pressure acquires an extremum which is a critical
point. In other words, this is the case in which phase transition at a
single point. For $q_{M-critical}<q_{M}$, pressure will be a decreasing
function of the horizon radius without any extremum, hence critical point.

It is possible to obtain the free energy of this case in following form

\begin{equation}
F=\frac{r_{+}^{d-3}}{16}k-\frac{\pi r_{+}^{d-1}}{\left( d-1\right) \left(
d-2\right) }P+\frac{\left( 2d-5\right) r_{+}^{d-3}}{8\left( d-2\right)
\left( d-3\right) }q_{M}^{2}+\frac{2\left( 2d-5\right) \pi -\left(
d-2\right) }{16\pi ^{3}\left( d-2\right) }\left( d-3\right) r_{+}^{d-3}\Phi
_{E}^{2}.
\end{equation}

The final subject of interest in this section is the heat capacity. It is a
matter of calculation to obtain the heat capacity as

\begin{equation}
C=\frac{\left( d-2\right) \left( d-3\right) \pi ^{3}r_{+}^{3d-2}k+16\pi
^{3}r_{+}^{3d}P-2\pi r_{+}^{d+4}q_{M}^{2}-2\left( d-3\right) ^{2}\pi
r_{+}^{3d-3}\Phi _{E}^{2}}{64\pi ^{3}r_{+}^{2d+2}P-4\pi ^{2}\left(
d-2\right) \left( d-3\right) r_{+}^{2d}k+8\pi ^{2}\left( 2d-5\right)
q_{M}^{2}r_{+}^{6}+8\left( d-3\right) ^{2}r_{+}^{2d}\Phi _{E}^{2}}\left(
d-2\right) ,
\end{equation}%
in which, for the $5$-dimensional case, one can obtain the corresponding
divergencies as

\begin{equation}
r_{+}(C(d=5)\longrightarrow \infty )=\frac{1}{12}\sqrt{\frac{\left( \frac{%
54\left( \pi ^{2}k-\frac{4\Phi _{E}^{2}}{3}\right) ^{2}}{A_{9}^{1/3}}%
+6A_{9}^{1/3}-24\Phi _{E}^{2}+18\pi ^{2}k\right) }{\pi ^{3}P}},
\end{equation}%
where $A_{9}=A_{10}+24\sqrt{15}\pi ^{4}Pq_{M}\sqrt{A_{11}}$, in which $%
A_{10} $ and $A_{11}$ are

\begin{eqnarray}
A_{10} &=&-4320\pi ^{8}P^{2}q_{M}^{2}+27\pi ^{6}k^{3}-108\pi ^{4}k^{2}\Phi
_{E}^{2}+144\pi ^{2}k\Phi _{E}^{4}-64\Phi _{E}^{6}, \\
&&  \notag \\
A_{11} &=&2160\pi ^{8}P^{2}q_{M}^{2}-27\pi ^{6}k^{3}+108\pi ^{4}k^{2}\Phi
_{E}^{2}-144\pi ^{2}k\Phi _{E}^{4}+64\Phi _{E}^{6}.
\end{eqnarray}

It is worthwhile to mention that the high energy limit and the asymptotic
behavior of the heat capacity for this case are, respectively, given by

\begin{eqnarray}
\lim_{r_{+}\rightarrow 0}C &=&-\frac{3\pi }{20}r_{+}^{3}+\frac{3}{25}\frac{%
3k\pi ^{2}-4\Phi _{E}^{2}}{\pi q_{M}^{2}}r_{+}^{7}+O\left( r_{+}^{9}\right) ,
\\
&&  \notag \\
\lim_{r_{+}\rightarrow \infty }C &=&\frac{3\pi }{4}r_{+}^{3}+\frac{3}{16}%
\frac{3k\pi ^{2}-4\Phi _{E}^{2}}{\pi ^{2}P}r_{+}+O\left( \frac{1}{r_{+}}%
\right) .
\end{eqnarray}

In order to understand more details regarding the types of phase transitions
of this case, we have plotted the necessary diagrams (see right panel of
Fig. \ref{FigE1}). First of all, in the absence of magnetic charge, it is
seen that only one divergency for the heat capacity exits. The sign of the
heat capacity changes from negative to positive as we cross such divergence
point. Therefore, we have a phase transition between smaller unstable black
holes to larger stable ones. In the presence of magnetic charge, depending
on the value of magnetic charge, heat capacity could enjoy: i) one root and
two divergencies in which stable black holes exist only between the root and
the smaller divergency, and also after larger divergency. ii) one root and
one divergency around which the sign of heat capacity is positive. In this
case, the phase transition is between two stable phases of smaller and
larger black holes. iii) one root in which after it, the heat capacity is
positive valued and solutions are thermally stable. Here, we see that
consideration of the constant electric field resulted in the presence of
peculiarities in thermodynamical behavior of the black holes. In the
presence of magnetic charge, the stability conditions, regions of stability
and absence of critical behavior depend on the value of magnetic charge (for
fixed values of other parameters). Once again, we point it out that in the
absence of electric field, it is possible to obtain critical behavior for
the magnetic black holes. In this case, the stability conditions, hence
place of the divergencies of heat capacity, are different but the end
results show the possibility of van der Waals like behavior and critical
points for magnetically charged black holes.

As for the high energy limit and asymptotic behavior of heat capacity,
evidently, these two limits are governed by the gravitational part of
solutions since the dominant terms for both cases depend only on the horizon
radius. The effects of the magnetic charge could be seen in the second
leading order term of the high energy limit in which the magnetic charge is
in the denominator of this term. Interestingly, for both cases of the
asymptotical behavior and high energy limit, the presence of electric field
is observed in the second leading order term in the numerator of both cases.
This confirms that the effects of magnetic charge on the thermal stability
of medium black holes become significant. It should be highlighted that in
the presence of constant electric field, the high energy limit and
asymptotic behavior of heat capacity were highly modified as compared to
other thermodynamical quantities. Once more, this highlights the differences
between these black holes and previous ones.


\section{Constant Magnetic field}

In this ensemble, we assume that the non-zero spatial component of the
electromagnetic tensor, $A_{\theta }$, is constant. Therefore, we replace
the magnetic charge, $q_{M},$ with the following relation

\begin{equation}
q_{M}=\frac{\left( d-3\right) r_{+}^{d-3}\Phi _{M}}{\pi },
\end{equation}%
in thermodynamical quantities. It is easy to show that the mass(enthalpy) of
this case is given by

\begin{equation}
M=\frac{\left( d-2\right) r_{+}^{d-3}}{16}k+\frac{\pi r_{+}^{d-1}}{d-1}P+%
\frac{r_{+}^{3-d}}{8\left( d-3\right) }q_{E}^{2}+\frac{\left( d-3\right)
r_{+}^{d-3}}{8\pi ^{2}}\Phi _{M}^{2}.
\end{equation}

Such as before, since the roots of $M$ could not be extracted analytically
for general $d$, we only present the roots for $5$-dimensional case

\begin{equation}
r_{+}(M(d=5)=0)=\frac{1}{6}\sqrt{\frac{\left( \frac{27\left( \pi ^{2}k+\frac{%
4\Phi _{M}^{2}}{3}\right) ^{2}}{A_{12}^{1/3}}+3A_{12}^{1/3}-12\Phi
_{M}^{2}-9\pi ^{2}k\right) }{\pi ^{3}P}},
\end{equation}%
where $A_{12}=A_{13}+12\sqrt{3}\pi ^{4}Pq_{E}\sqrt{A_{14}}$, in which $%
A_{13} $ and $A_{14}$ are in the following forms

\begin{eqnarray}
A_{13} &=&-216\pi ^{8}P^{2}q_{E}^{2}-27\pi ^{6}k^{3}-108\pi ^{4}k^{2}\Phi
_{M}^{2}-144\pi ^{2}k\Phi _{M}^{4}-64\Phi _{M}^{6}, \\
&&  \notag \\
A_{14} &=&108\pi ^{8}P^{2}q_{E}^{2}+27\pi ^{6}k^{3}+108\pi ^{4}k^{2}\Phi
_{M}^{2}+144\pi ^{2}k\Phi _{M}^{4}+64\Phi _{M}^{6}.
\end{eqnarray}

The high energy limit of the mass/enthalpy and its asymptotic behavior are,
respectively, as follows

\begin{eqnarray}
\lim_{r_{+}\rightarrow 0}M &=&\frac{1}{16r_{+}^{2}}q_{E}^{2}+\left( \frac{3}{%
16}k+\frac{1}{4\pi ^{2}}\Phi _{M}^{2}\right) r_{+}^{2}+O\left(
r_{+}^{4}\right) , \\
&&  \notag \\
\lim_{r_{+}\rightarrow \infty }M &=&\frac{\pi }{4}r_{+}^{4}P+\left( \frac{3}{%
16}k+\frac{1}{4\pi ^{2}}\Phi _{M}^{2}\right) r_{+}^{2}+O\left( \frac{1}{%
r_{+}^{2}}\right) .
\end{eqnarray}

For the high energy limit, contrary to the previous cases, the dominant term
includes a coupling between electric charge and horizon radius. Here, the
effects of the magnetic field could be detected in the second dominant term
of high energy limit. Interestingly, for

\begin{equation}
k=-\frac{4\Phi _{M}^{2}}{3\pi ^{2}},
\end{equation}%
the effects of topological structure and magnetic potential on mass/enthalpy
cancel out. It is evident that such a case, only happens for black holes
with hyperbolic horizon. By studying the asymptotic behavior, one can
observe that the first dominant term includes the pressure while the second
dominant term in this limit is the same as second dominant term in the high
energy limit. This property enables us to categorize the mass/enthalpy
behavior of black holes into three categories: i) small black holes in which
the general behavior significantly depends on the electric charge. ii)
medium black holes which are governed by the magnetic potential and
topological structure of black holes. iii) large black holes which are
mainly described by the pressure, hence the cosmological constant. The
existence of root for the mass/enthalpy depends on the following condition

\begin{equation}
k<-\frac{4\Phi _{M}^{2}}{3\pi ^{2}}.
\end{equation}

\begin{figure}[tbp]
$%
\begin{array}{ccc}
\epsfxsize=5.75cm \epsffile{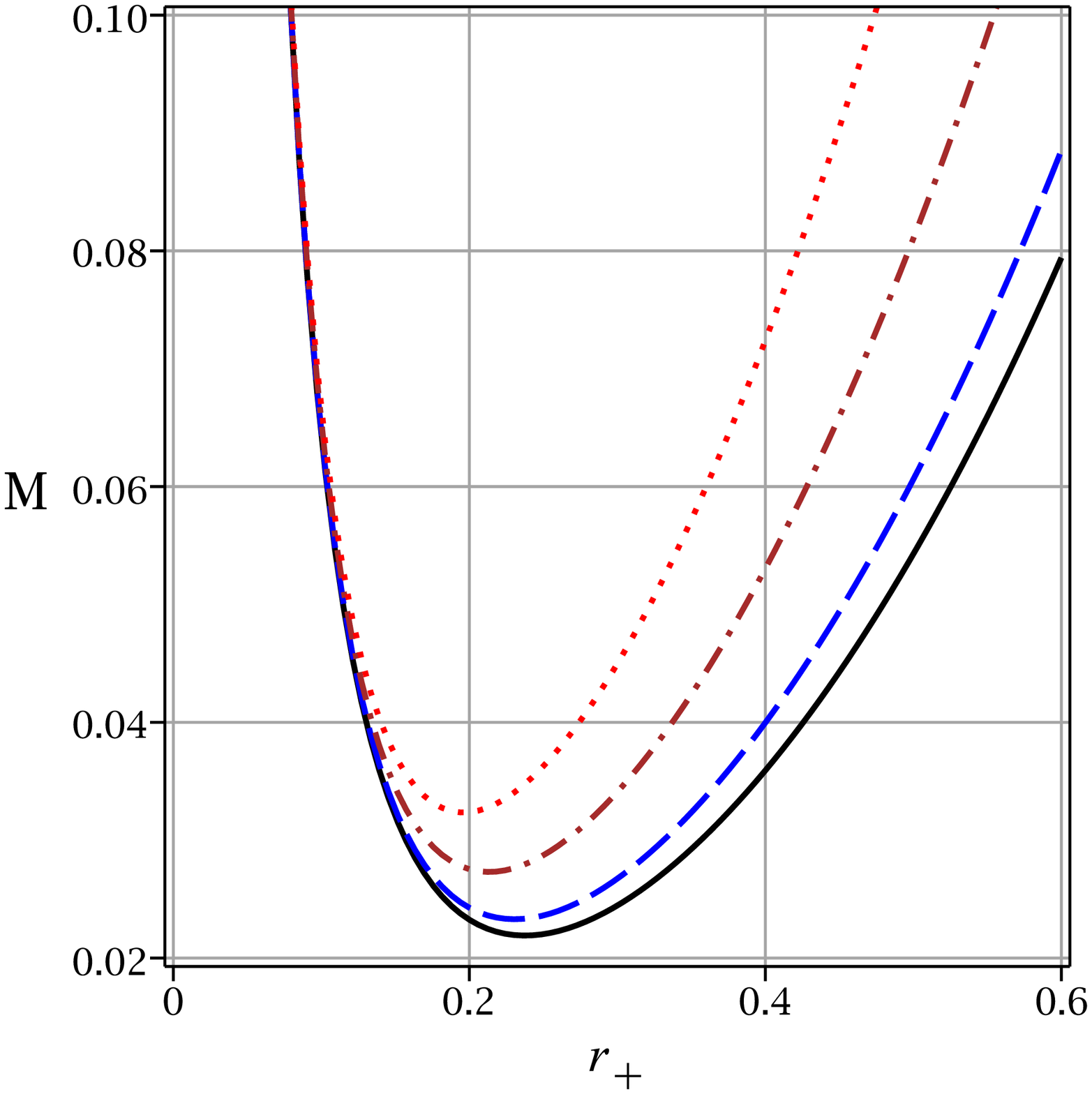} & \epsfxsize=5.75cm \epsffile{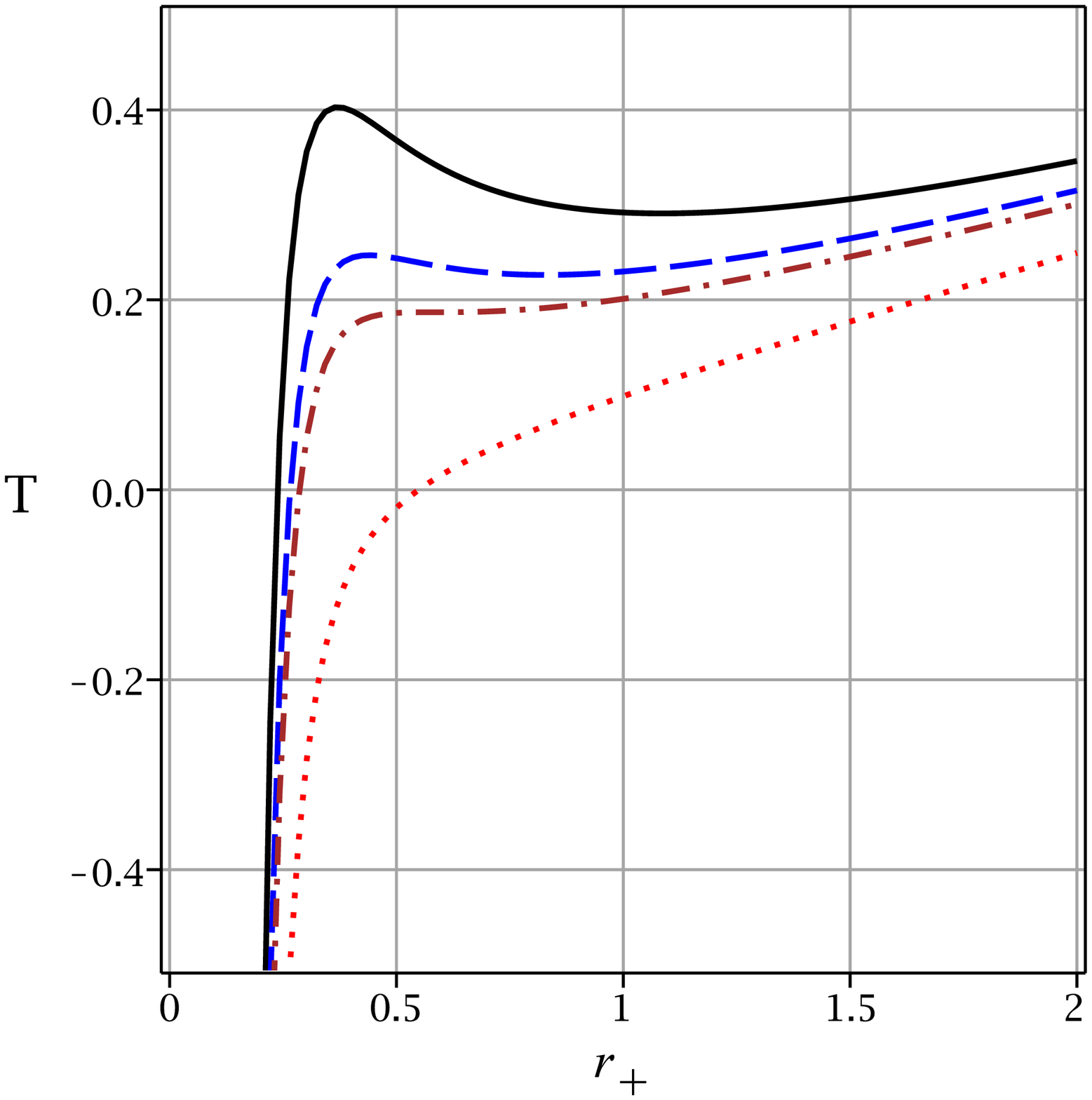}
& \epsfxsize=5.75cm \epsffile{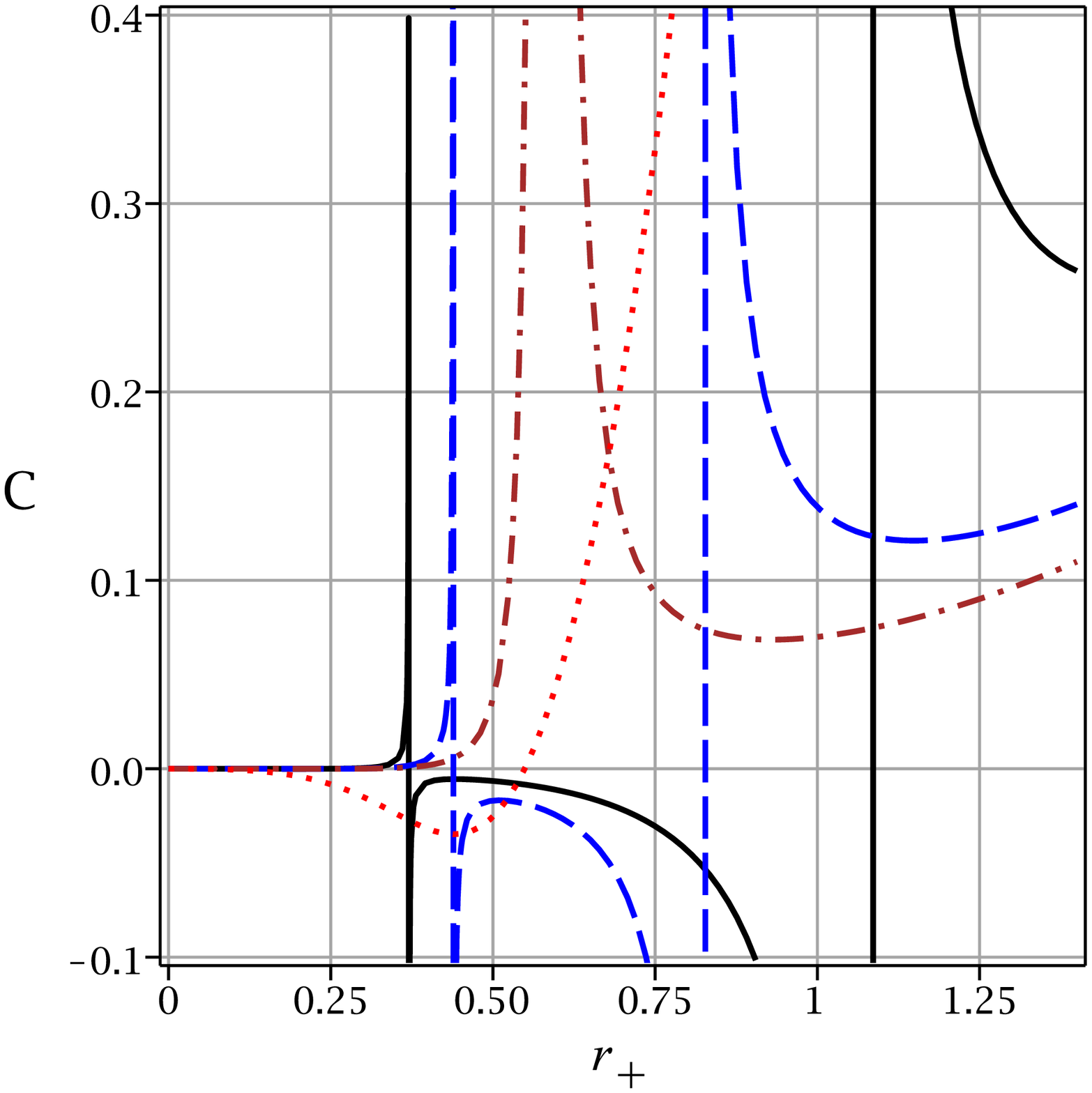}%
\end{array}
$%
\caption{$H$ (left panel), $T$ (middle panel) and $C$ (right panel) versus $%
r_{+}$ for $d=5$, $k=1$, $P=0.1$ and $q _{E}=0.1$; $\Phi_{M}=0$ (continuous
line), $\Phi_{M}=1.7$ (dashed line), $\Phi_{M}=2.056$ (dashed-dotted line)
and $\Phi_{M}=3$ (dotted line).}
\label{FigM1}
\end{figure}

If this condition is satisfied, mass/enthalpy of the black holes could
acquire root. Such case could only happen for black holes with hyperbolic
horizon. In order to have a better picture regarding the behavior of the
mass/enthalpy, we have plotted them in left panel of Fig. \ref{FigM1}. The
mass/enthalpy for this case has a minimum which is an increasing function of
the magnetic potential.

The temperature for this case is calculated as

\begin{equation}
T=\frac{1}{2\pi }\left[ \frac{\left( d-3\right) }{2r_{+}}k+\frac{8\pi r_{+}}{%
\left( d-2\right) }P-\frac{q_{E}^{2}}{\left( d-2\right) r_{+}^{2d-5}}-\frac{%
\left( d-3\right) ^{2}}{\pi ^{2}\left( d-2\right) r_{+}}\Phi _{M}^{2}\right]
.
\end{equation}

Such as before, it is not possible to extract the roots of temperature
analytically for the general $d$-dimensional case, but in $5$-dimension, we
have the following root

\begin{equation}
r(T(d=5)=0)=\frac{1}{12}\sqrt{\frac{\left( \frac{54\left( \pi ^{2}k-\frac{%
4\Phi _{M}^{2}}{3}\right) ^{2}}{A_{15}^{1/3}}+6A_{15}^{1/3}+24\Phi
_{M}^{2}-18\pi ^{2}k\right) }{\pi ^{3}P}},
\end{equation}%
where $A_{15}=A_{16}+24\sqrt{3}\pi ^{4}Pq_{E}\sqrt{A_{17}}$, in which $%
A_{16} $ and $A_{17}$ are

\begin{eqnarray}
A_{16} &=&864\pi ^{8}P^{2}q_{E}^{2}-27\pi ^{6}k^{3}+108\pi ^{4}k^{2}\Phi
_{M}^{2}-144\pi ^{2}k\Phi _{M}^{4}+64\Phi _{M}^{6}, \\
&&  \notag \\
A_{17} &=&432\pi ^{8}P^{2}q_{E}^{2}-27\pi ^{6}k^{3}+108\pi ^{4}k^{2}\Phi
_{M}^{2}-144\pi ^{2}k\Phi _{M}^{4}+64\Phi _{M}^{6}.
\end{eqnarray}

The high energy limit and asymptotic behavior of the temperature, for this
case could be written as

\begin{eqnarray}
\lim_{r_{+}\rightarrow 0}T &=&-\frac{1}{6\pi r_{+}^{5}}q_{E}^{2}+\left(
\frac{1}{2\pi }k-\frac{2}{3\pi ^{3}}\Phi _{M}^{2}\right) \frac{1}{r_{+}}%
+O\left( r_{+}\right) , \\
&&  \notag \\
\lim_{r_{+}\rightarrow \infty }T &=&\frac{4}{3}r_{+}P+\left( \frac{1}{2\pi }%
k-\frac{2}{3\pi ^{3}}\Phi _{M}^{2}\right) \frac{1}{r_{+}}+O\left( \frac{1}{%
r_{+}^{5}}\right) .
\end{eqnarray}

Here too, similar to the previous case, the dominant term of high energy
limit is governed by the electric charge. But contrary to mass/enthalpy, the
sign of the dominant term for high energy limit is negative. The second
dominant term includes both topological factor and magnetic potential. It is
possible to cancel out the effects of topological factor and magnetic
potential by following adjustment

\begin{equation}
k=\frac{4\Phi _{M}^{2}}{3\pi ^{2}},
\end{equation}%
which takes place only for spherical black holes. Considering that the
dominant term in the asymptotic behavior is positive valued, one can
conclude that at least one root is available for the temperature of these
black holes. In order to have a better picture regarding the effects of
magnetic potential on the behavior of temperature, we have plotted the
corresponding diagrams in the middle panel of Fig. \ref{FigM1}. Evidently,
the place of root and existence of extremum for the temperature and its
number are functions of the magnetic potential. The root of temperature is
an increasing function of the magnetic potential while the number of extrema
is a decreasing function of it. There exists a critical magnetic potential, $%
\Phi _{M-critical}$, which for magnetic potentials less than it, temperature
has two extrema. In addition, the distance between these extrema is a
decreasing function of the magnetic potential. When $\Phi _{M}$ reaches $%
\Phi _{M-critical}$, there will be only one extremum. Beyond this magnetic
potential, temperature will be an increasing function of the horizon radius
with one root and without extremum. Remembering that extremum in temperature
is where the heat capacity acquires divergency (thermal phase transition),
one can conclude that by increasing the magnetic potential to certain
values, the black hole will not have any thermal phase transition.
Therefore, for fixing other quantities and varying the magnetic potential,
one can state that for super magnetized black holes, no phase transition
exists. It is worthwhile to mention that the phase transition observed in
temperature versus horizon radius diagrams is van der Waals like. The nature
of the phase transition will be discussed when we turn to the heat capacity.

Using the temperature, it is a matter of calculation to obtain the following
relation for the pressure

\begin{equation}
P=\frac{1}{2\pi }\left[ \frac{\pi \left( d-2\right) }{2r_{+}}T-\frac{\left(
d-2\right) \left( d-3\right) }{8r_{+}^{2}}k+\frac{q_{E}^{2}}{4r_{+}^{2d-4}}+%
\frac{\left( d-3\right) ^{2}}{4\pi ^{2}r_{+}^{2}}\Phi _{M}^{2}\right] .
\end{equation}

The following relation can be obtained for the critical horizon radius

\begin{equation}
\frac{\left( d-2\right) \left( 4d-10\right) }{r_{+}^{2d-6}}\pi
^{2}q_{E}^{2}-\left( d-3\right) \left[ k\left( d-2\right) \pi ^{2}-2\left(
d-3\right) \Phi _{M}^{2}\right] =0,
\end{equation}%
which yields the following result

\begin{equation}
r_{c}=\left[ \frac{\left( d-3\right) \left[ k\left( d-2\right) \pi
^{2}-2\left( d-3\right) \Phi _{M}^{2}\right] }{2\left( d-2\right) \left(
2d-5\right) \pi ^{2}q_{E}^{2}}\right] ^{\frac{1}{2d-6}}.
\end{equation}

Using this critical horizon radius, one can extract critical temperature and
pressure in the following forms

\begin{eqnarray}
T_{c} &=&\left( 2d-4\right) ^{\frac{7-2d}{2d-6}}\left[ \left( 2d-5\right)
q_{E}^{2}\right] ^{\frac{1}{2d-6}}\pi ^{\frac{10-3d}{d-3}}\left[ \left(
d-3\right) \left( d-2\right) \pi ^{2}k-2\left( d-3\right) ^{2}\Phi _{M}^{2}%
\right] ^{\frac{5-2d}{2d-6}}  \notag \\
&&  \notag \\
&&\left[ \pi ^{4}\left( d-2\right) ^{2}\left( d-3\right) ^{2}k^{2}-4\left(
d-2\right) \left( d-3\right) ^{3}\pi ^{2}\Phi _{M}^{2}k-4\pi ^{4}\left(
d-2\right) ^{2}\left( 2d-5\right) q_{E}^{4}+4\left( d-3\right) ^{4}\Phi
_{M}^{4}\right] ,
\end{eqnarray}

\begin{eqnarray}
P_{c} &=&\frac{\left( \Phi _{M}^{2}\left( 6-2d\right) +\left( d-2\right) \pi
^{2}k\right) ^{\frac{2-d}{d-3}}}{4\left( d-3\right) ^{\frac{d-2}{d-3}}}%
\left\{ q_{E}^{\frac{2d-4}{d-3}}\left( q_{E}^{2}+q_{M}^{2}\right) \left[
2\pi ^{d-1}\left( d-2\right) ^{d-2}\left( 2d-5\right) ^{d-2}\right]
^{1/\left( d-3\right) }\right.  \notag \\
&&  \notag \\
&&\left. +\left[ \left( d-2\right) \left( 2d-5\right) \right] ^{1/\left(
d-3\right) }\left[ 2^{\frac{4-d}{d-3}}\Gamma ^{\prime }-20d^{3}\pi ^{\frac{%
d-1}{d-3}}\left( k^{2}\left( \frac{q_{E}^{2}}{2^{2d-7}}\right) ^{1/\left(
d-3\right) }+\frac{\left( 2q_{E}^{4d-10}\right) ^{1/\left( d-3\right) }}{5}%
\right) \right] \right\}  \notag \\
&&  \notag \\
&&-\frac{1}{16}\left( \frac{2\pi ^{5-d}\left( d-2\right) ^{d-2}\left(
d-3\right) ^{d-4}\left( 2d-5\right) kq_{E}^{2}}{\Phi _{M}^{2}\left(
6-2d\right) +\pi ^{2}k\left( d-2\right) }\right) ^{1/\left( d-3\right) },
\end{eqnarray}%
in which

\begin{eqnarray*}
\Gamma ^{\prime } &=&52\left( \frac{13d^{2}-28d+20}{13}\right) \left( \pi
^{d-1}q_{E}^{4d-10}\right) ^{1/\left( d-3\right) }-4k\Phi _{M}^{2}\left(
d-2\right) \left( d-3\right) ^{3}\left( \frac{q_{E}^{2}}{\pi ^{d-5}}\right)
^{1/\left( d-3\right) } \\
&&+\left( \frac{4\left( d-3\right) ^{4}\Phi _{M}^{2}}{\pi ^{\frac{3d-11}{d-3}%
}}+\pi ^{\frac{d-1}{d-3}}k^{2}\left( d^{4}+37d^{2}-60d+36\right) \right)
q_{E}^{\frac{4d-10}{d-3}}
\end{eqnarray*}

\begin{figure}[tbp]
$%
\begin{array}{c}
\epsfxsize=7cm \epsffile{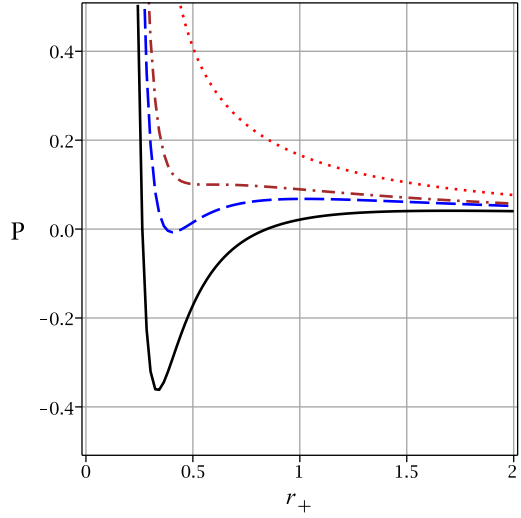}%
\end{array}
$%
\caption{$P$ versus $r_{+}$ for $d=5$, $k=1$, $T=0.187$ and $q_{E}=0.1$; $%
\Phi_{M}=0$ (continuous line), $\Phi_{M}=1.7$ (dashed line), $\Phi_{M}=2.056$
(dashed-dotted line) and $\Phi_{M}=3$ (dotted line).}
\label{FigM2}
\end{figure}

The plotted diagram for the pressure (Fig. \ref{FigM2}) confirms that
depending on choices for the magnetic potential, the solutions would have a
van der Waals like phase transition. Interestingly, the critical horizon
radius obtained for this case has fundamental differences comparing to the
previous case in which electric field was constant. Here, the presence of
magnetic potential is present in the numerator of obtained critical horizon
radius while the electric charge is in the denominator. This is opposite to
what was observed for critical horizon radius in case of constant electric
field. It is worthwhile to mention that critical horizon radius is a
decreasing function of magnetic potential and electric charge, whereas the
critical temperature and pressure are increasing functions of them.

It is possible to obtain the free energy of this case as

\begin{equation}
F=\frac{r_{+}^{d-3}}{16}k-\frac{\pi r_{+}^{d-1}}{\left( d-1\right) \left(
d-2\right) }P+\frac{\left( 2d-5\right) \left( d-3\right) r_{+}^{d-3}}{8\pi
^{2}\left( d-2\right) }\Phi _{M}^{2}+\frac{2\left( 2d-5\right) \pi -\left(
d-2\right) }{16\pi \left( d-2\right) \left( d-3\right) }r_{+}^{3-d}q_{E}^{2}.
\end{equation}

Our final remark in this section concerns the thermal stability in the
context of canonical ensemble, hence heat capacity. It is a matter of
calculation to obtain the heat capacity as

\begin{equation}
C=\frac{\left( d-2\right) \left( d-3\right) \pi ^{3}r_{+}^{3d-2}k+16\pi
^{4}r_{+}^{3d}P-2\pi ^{3}r_{+}^{d+4}q_{E}^{2}-2\left( d-3\right) ^{2}\pi
r_{+}^{3d-2}\Phi _{M}^{2}}{64\pi ^{3}r_{+}^{2d+2}P-4\pi ^{2}\left(
d-2\right) \left( d-3\right) r_{+}^{2d}k+8\pi ^{2}\left( 2d-5\right)
q_{E}^{2}r_{+}^{6}+8\left( d-3\right) ^{2}r_{+}^{2d}\Phi _{M}^{2}}\left(
d-2\right) .
\end{equation}

The divergencies of heat capacity for $d$-dimensional case could not be
obtained analytically, but for the $5$-dimensional case, the divergencies
are given by

\begin{equation}
r(C(d=5)\longrightarrow \infty )=\frac{1}{12}\sqrt{\frac{\left( \frac{%
54\left( \pi ^{2}k-\frac{4\Phi _{M}^{2}}{3}\right) ^{2}}{A_{18}^{1/3}}%
+6A_{18}^{1/3}-24\Phi _{M}^{2}+18\pi ^{2}k\right) }{\pi ^{3}P}},
\end{equation}%
where $A_{18}=A_{19}+24\sqrt{15}\pi ^{4}Pq_{E}\sqrt{A_{20}}$, in which $%
A_{19}$ and $A_{20}$ are in the following forms

\begin{eqnarray}
A_{19} &=&-4320\pi ^{8}P^{2}q_{E}^{2}+27\pi ^{6}k^{3}-108\pi ^{4}k^{2}\Phi
_{M}^{2}+144\pi ^{2}k\Phi _{M}^{4}-64\Phi _{M}^{6}, \\
&&  \notag \\
A_{20} &=&2160\pi ^{8}P^{2}q_{E}^{2}-27\pi ^{6}k^{3}+108\pi ^{4}k^{2}\Phi
_{M}^{2}-144\pi ^{2}k\Phi _{M}^{4}+64\Phi _{M}^{6}.
\end{eqnarray}

It is possible to obtain the high energy limit and asymptotic behavior of
the heat capacity for this case in the following forms

\begin{eqnarray}
\lim_{r_{+}\rightarrow 0}C &=&-\frac{3\pi }{20}r_{+}^{3}+\frac{3}{25}\frac{%
3k\pi ^{2}-4\Phi _{M}^{2}}{\pi q_{E}^{2}}r_{+}^{7}+O\left( r_{+}^{9}\right) ,
\\
&&  \notag \\
\lim_{r_{+}\rightarrow \infty }C &=&\frac{3\pi }{4}r_{+}^{3}+\frac{3}{16}%
\frac{3k\pi ^{2}-4\Phi _{M}^{2}}{\pi ^{2}P}r_{+}+O\left( \frac{1}{r_{+}}%
\right) .
\end{eqnarray}

First of all, the dominant terms in the high energy limit and asymptotic
behavior depend only on the horizon radius which confirms that these two
terms are governed by the gravitational sector of the action. On the other
hand, the second dominant term of high energy limit depends on topological
factor, magnetic potential and electric charge. This term is an increasing
function of the topological term while it is a decreasing function of the
magnetic potential and electric charge. The presence of electric charge
could be seen in the denominator of the second dominant term. As for the
asymptotical behavior, the effects of pressure could be observed in the
second dominant term of it. The pressure is in denominator of this term. In
order to understand the behavior of the heat capacity in more details, one
may refer to right panel of Fig. \ref{FigM1}. In the case of $\Phi_{M}=\Phi
_{M-critical}$, the heat capacity enjoys the presence of a root and
divergency which is located after the root. Around the divergency, the sign
of heat capacity is positive. Therefore, here we have a phase transition
between two stable black holes. On the other hand, for $\Phi _{M}<\Phi
_{M-critical}$, the heat capacity enjoys a root and two divergencies. The
divergencies are located after the root, and between them, the heat capacity
is negative. In other words, between the root and smaller divergency, and
after larger divergency, the heat capacity is positive. This indicates that
over a specific region (between divergency), black holes suffer from
instability. But comparing this case with pressure or temperature diagrams,
one can see that no-physical black holes exists for this region. Therefore,
there is a phase transition over a region (between the divergencies) for
black holes in this case. By increasing the magnetic potential beyond $\Phi
_{M-critical}$, \ the divergencies in the heat capacity are eliminated and
black holes will have no thermal phase transition in their structure. It is
worthwhile to mention that before root, both temperature and heat capacity
are negative valued which indicates that solutions in this region are not
physical ones (according to classical concepts of thermodynamics of black
holes). This case takes place for all mentioned cases of the magnetic
potential.


\section{Constant magnetic and electric fields}

In this section, we assume that the temporal and spatial components of the
electromagnetic tensor, $A_{t}$ and $A_{\theta }$, are constant. So, we
replace the electric and magnetic charges, $q_{E}$ and $q_{M},$ with the
following relations

\begin{equation}
q_{M}=\frac{\left( d-3\right) r_{+}^{d-3}\Phi _{M}}{\pi },~~~\&~~~q_{E}=%
\frac{\left( d-3\right) r_{+}^{d-3}\Phi _{E}}{\pi }.
\end{equation}%
in thermodynamical quantities. Therefore, the mass/enthalpy of this
consideration can be rewritten as

\begin{equation}
M=\frac{\left( d-2\right) r_{+}^{d-3}}{16}k+\frac{\pi r_{+}^{d-1}}{d-1}P+%
\frac{\left( d-3\right) r_{+}^{d-3}}{8\pi ^{2}}\left( \Phi _{M}^{2}+\Phi
_{E}^{2}\right) .
\end{equation}%
for which, the roots are obtained, analytically, as

\begin{equation}
r_{+}(M=0)=\left( 4\pi ^{\frac{3}{2}}\sqrt{-\frac{P}{\left( d-1\right) \left[
\pi ^{2}\left( d-2\right) k+2\left( d-3\right) \left( \Phi _{M}^{2}+\Phi
_{E}^{2}\right) \right] }}\right) ^{-1}.
\end{equation}

Evidently, in order to have a root for the mass/enthalpy, the following
inequality should be satisfied

\begin{equation}
k<-\frac{2\left( d-3\right) \left( \Phi _{M}^{2}+\Phi _{E}^{2}\right) }{\pi
^{2}\left( d-2\right) }.
\end{equation}

This relation could be satisfied for black holes with hyperbolic horizon. In
other words, the mass/enthalpy enjoys the absence of root in the cases of
black holes with flat or spherical horizons. It is worthwhile to mention
that the root of mass is an increasing function of electric and magnetic
potentials while it is a decreasing function of pressure. The high energy
and asymptotic behavior of the mass/enthalpy for $5$-dimensional case are
given by

\begin{eqnarray}
\lim_{r_{+}\rightarrow 0}M &=&\left( \frac{3}{16}k+\frac{1}{4\pi ^{2}}\left[
\Phi _{M}^{2}+\Phi _{E}^{2}\right] \right) r_{+}^{2}+\frac{\pi }{4}%
r_{+}^{4}P, \\
&&  \notag \\
\lim_{r_{+}\rightarrow \infty }M &=&\frac{\pi }{4}r_{+}^{4}P+\left( \frac{3}{%
16}k+\frac{1}{4\pi ^{2}}\left[ \Phi _{M}^{2}+\Phi _{E}^{2}\right] \right)
r_{+}^{2}
\end{eqnarray}

\begin{figure}[tbp]
$%
\begin{array}{ccc}
\epsfxsize=5.75cm \epsffile{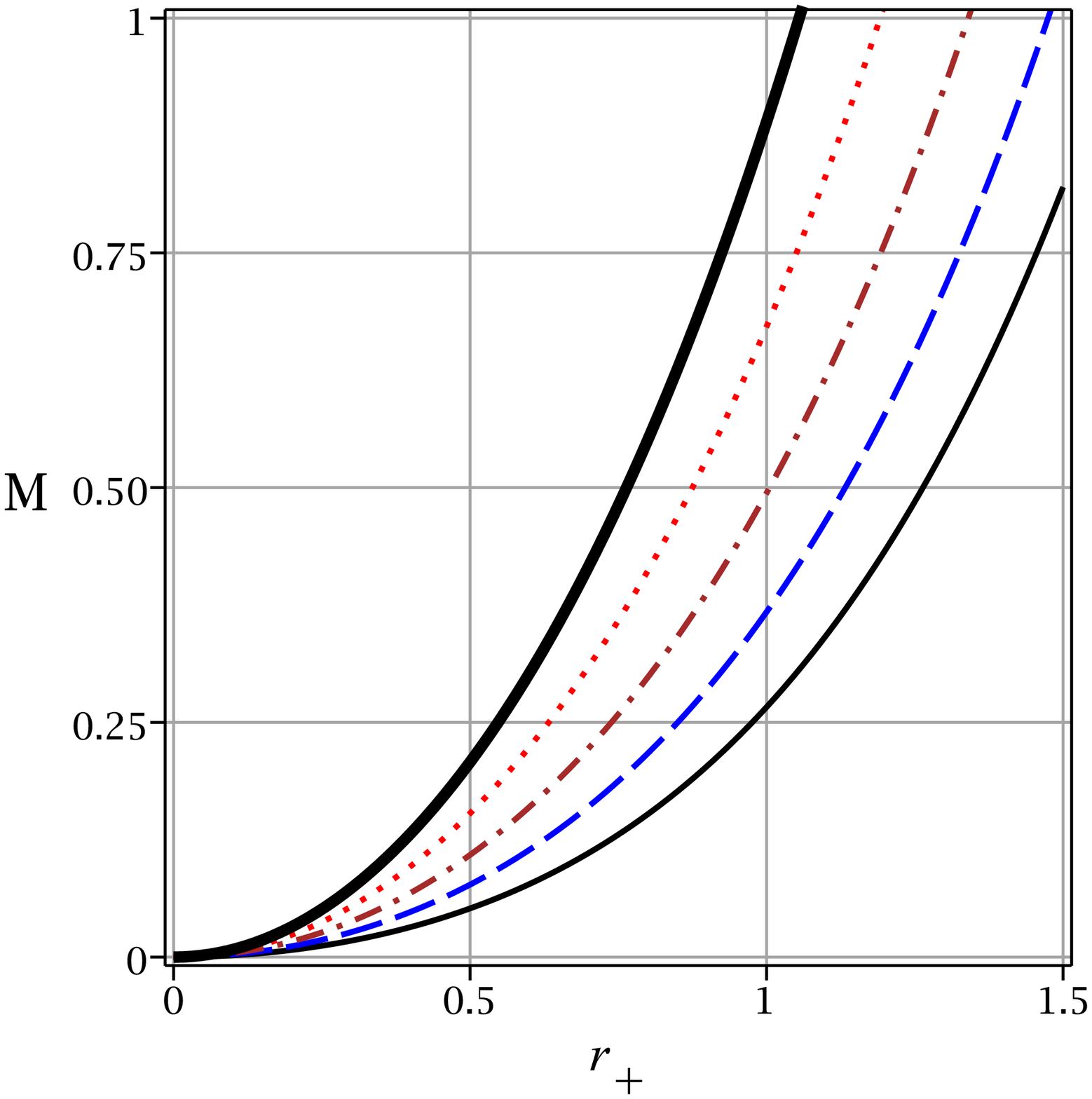} & \epsfxsize=5.75cm \epsffile{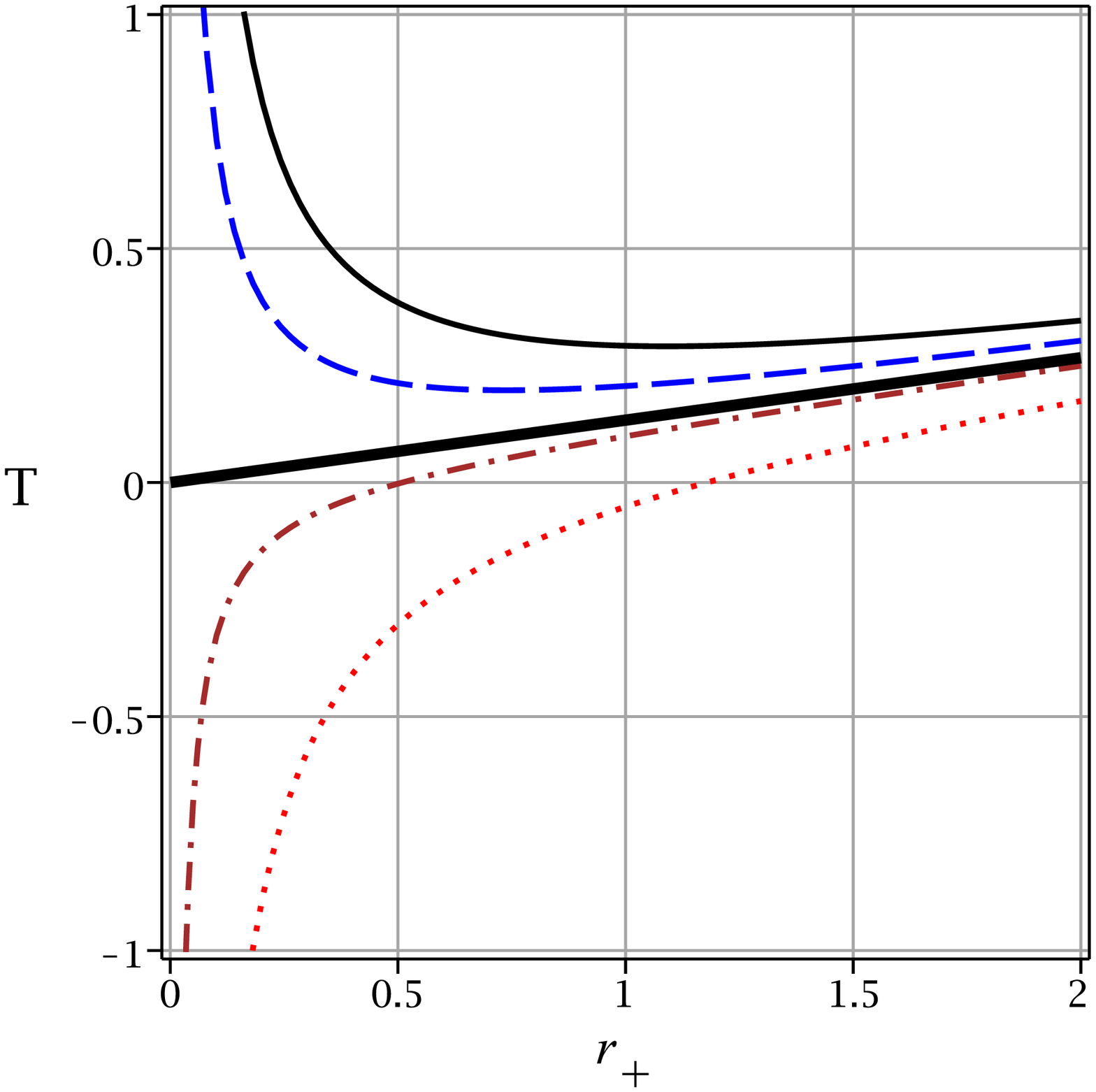}
& \epsfxsize=5.75cm \epsffile{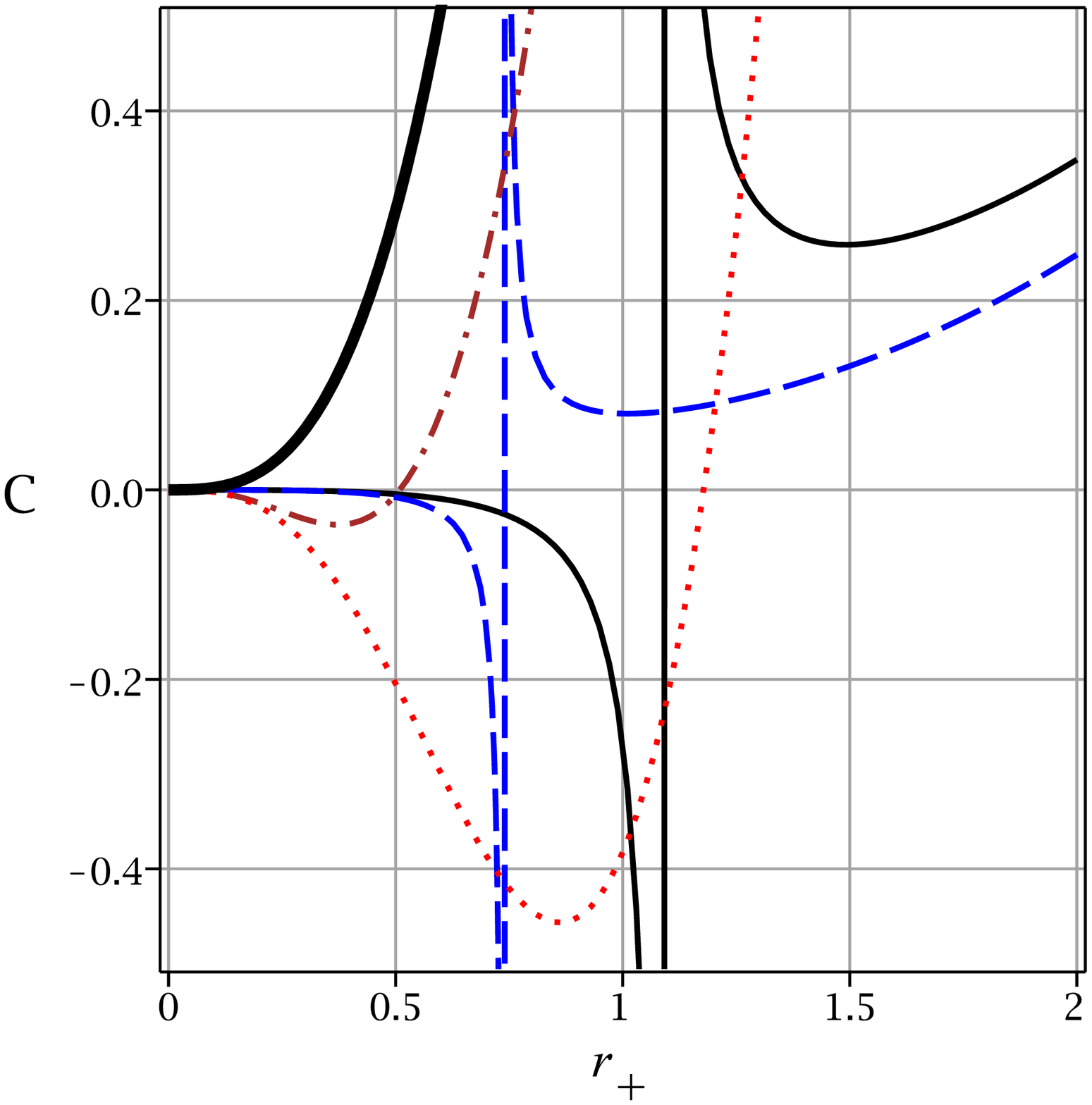}%
\end{array}
$%
\caption{$M$ (left panel), $T$ (middle panel) and $C$ (right panel) versus $%
r_{+}$ for $d=5$, $P=0.1$ and $\Phi _{E}=0.1$; $k=1$, $\Phi _{M}=0$
(continuous line), $\Phi _{M}=2$ (dashed line), $\Phi _{M}=3$ (dashed-dotted
line) and $\Phi _{M}=4$ (dotted line). \textbf{The bold line:} $k=\frac{%
2\left( d-3\right) \left( \Phi _{M}^{2}+\Phi _{E}^{2}\right) }{\protect\pi %
^{2}\left( d-2\right) }$ and $\Phi _{M}=4$.}
\label{FigEM1}
\end{figure}

Here, the dominant term in high energy limit includes electric and magnetic
potential with topological factors. Comparing this with previous cases, one
can notice that for vanishing horizon radius, mass also vanishes which is
opposite to previous cases. Such a behavior is rooted in consideration of
the constant electric and magnetic fields. This is, in the absence of
horizon radius, the mass/enthalpy does diverge. Instead it tends to a
constant (zero) (see left panel of Fig. \ref{FigEM1}). On the other hand,
the asymptotic behavior is governed by the pressure term. This confirms that
highest modification for considering constant electric and magnetic fields
could be observed for small black holes only. Careful examination of these
limits also confirms one more result: with such consideration, one can see
that black holes are separated into only two groups of small and large black
holes. In other words, the class of the medium black holes for the case
under consideration is eliminated. It is worth to mention that for the
choice of

\begin{equation}
k=-\frac{2\left( d-3\right) \left( \Phi _{M}^{2}+\Phi _{E}^{2}\right) }{\pi
^{2}\left( d-2\right) },
\end{equation}%
the effects of the electric and magnetic charges, and topological terms of
the black hole are canceled out and the behavior of mass/enthalpy of the
black holes, irrespective of their size, is determined by the pressure.

The temperature for this case is

\begin{equation}
T=\frac{1}{2\pi }\left[ \frac{\left( d-3\right) }{2r_{+}}k+\frac{8\pi r_{+}}{%
\left( d-2\right) }P-\frac{\left( d-3\right) ^{2}}{\pi ^{2}\left( d-2\right)
r_{+}}\left( \Phi _{M}^{2}+\Phi _{E}^{2}\right) \right] .
\end{equation}

The root of the temperature for this case is obtained as

\begin{equation}
r_{+}(T=0)=\frac{1}{4P\pi ^{\frac{3}{2}}}\sqrt{P\left( d-1\right) \left[
2\left( d-3\right) \left( \Phi _{M}^{2}+\Phi _{E}^{2}\right) -\pi ^{2}\left(
d-2\right) k\right] }.
\end{equation}

First of all, the root of temperature is a decreasing function of the
pressure and topological factor while it is an increasing function of the
electric and magnetic potentials. The existence of root for the temperature
depends on satisfaction of the following condition

\begin{equation}
\frac{2\left( d-3\right) \left( \Phi _{M}^{2}+\Phi _{E}^{2}\right) }{\pi
^{2}\left( d-2\right) }>k.
\end{equation}

The mentioned condition is automatically satisfied for black holes with flat
and hyperbolic horizons, whereas for the spherical one, as long as $2\left(
\Phi _{M}^{2}+\Phi _{E}^{2}\right) \geq \pi ^{2}$, the mentioned condition
would be satisfied, hence spherical black holes would have root in their
temperature. It is worth mentioning that for

\begin{equation}
k=\frac{2\left( d-3\right) }{\pi ^{2}\left( d-2\right) }\left( \Phi
_{M}^{2}+\Phi _{E}^{2}\right) ,
\end{equation}%
the effects of electric and magnetic potentials, and topological term of the
black holes on the temperature are canceled and the behavior of temperature
is only determined by the pressure which will be only an increasing function
of horizon radius. The high energy limit and asymptotic behavior of the
solutions for $5$-dimensional black holes are given by

\begin{eqnarray}
\lim_{r_{+}\rightarrow 0}T &=&\left( \frac{1}{2\pi }k-\frac{2}{3\pi ^{3}}%
\left[ \Phi _{M}^{2}+\Phi _{E}^{2}\right] \right) \frac{1}{r_{+}}+\frac{4}{3}%
Pr_{+}, \\
&&  \notag \\
\lim_{r_{+}\rightarrow \infty }T &=&\frac{4}{3}Pr_{+}+\left( \frac{1}{2\pi }%
k-\frac{2}{3\pi ^{3}}\left[ \Phi _{M}^{2}+\Phi _{E}^{2}\right] \right) \frac{%
1}{r_{+}}.
\end{eqnarray}

The high energy limit of temperature, similar to mass, is governed by
electric and magnetic potentials, and topological factor. If, $\frac{
2\left( d-3\right) \left( \Phi _{M}^{2}+\Phi _{E}^{2}\right) }{\pi
^{2}\left( d-2\right) }=k$, then the temperature of black holes for
vanishing horizon radius, vanishes too. In other words, in this case, for
vanishing horizon radius, temperature will have finite (zero) value. On the
other hand, if $\frac{2\left( d-3\right) \left( \Phi _{M}^{2}+\Phi
_{E}^{2}\right) }{\pi ^{2}\left( d-2\right) }=k$ is not satisfied, the
temperature for $r_{+}\rightarrow 0$ will diverge. The possibility of zero
temperature for $r_{+}\rightarrow 0$ resulted from the constant electric and
magnetic fields. In the absence of electric field, this property is still
valid. The absence of divergent temperature for $r_{+}\rightarrow 0$ is
observed for specific black holes. Here, we see that by modification in the
action and with special consideration for matter field, this behavior
happens. In order to have a better picture regrading the behavior of
temperature, we have plotted a diagram (see middle panel of Fig. \ref{FigEM1}%
). Evidently, the behavior of the temperature could be divided into three
classes: i) if $\frac{2\left( d-3\right) \left( \Phi _{M}^{2}+\Phi
_{E}^{2}\right) }{\pi ^{2}\left( d-2\right) }>k$, the temperature is an
increasing function of the horizon radius with one root. Before the root,
temperature is negative valued and solutions are not physical. ii) if $\frac{%
2\left( d-3\right) \left( \Phi _{M}^{2}+\Phi _{E}^{2}\right) }{\pi
^{2}\left( d-2\right) }=k$, the effects of the electric and magnetic
potentials, and topological factor are canceled out and temperature will be
positive valued everywhere and only an increasing function of the horizon
radius. Finally for the case $\frac{2\left( d-3\right) \left( \Phi
_{M}^{2}+\Phi _{E}^{2}\right) }{\pi ^{2}\left( d-2\right) }<k$, the
temperature will be positive valued with one minimum. The minimum is a
decreasing function of the magnetic and electric potentials. Since the
presence of extremum in temperature indicates the existence of divergency
(phase transition) in the heat capacity, one can conclude that only for the
third case (for $\frac{2\left( d-3\right) \left( \Phi _{M}^{2}+\Phi
_{E}^{2}\right) }{\pi ^{2}\left( d-2\right) }<k$) solutions could undergo
phase transition like behavior. We will examine the type of this phase
transition in the context of heat capacity later.

Using the obtained temperature, it is possible to have pressure in the
following form

\begin{equation}
P=\frac{1}{2\pi }\left[ \frac{\pi \left( d-2\right) }{2r_{+}}T-\frac{\left(
d-2\right) \left( d-3\right) }{8r_{+}^{2}}k+\frac{\left( d-3\right) ^{2}}{%
4\pi ^{2}r_{+}^{2}}\left( \Phi _{M}^{2}+\Phi _{E}^{2}\right) \right] .
\end{equation}

Now, using the properties of inflection point, one can find the following
relation governing critical point

\begin{equation}
\left( d-3\right) \left[ k\left( d-2\right) \pi ^{2}-2\left( d-3\right)
\left( \Phi _{M}^{2}+\Phi _{E}^{2}\right) \right] =0.
\end{equation}

\begin{figure}[tbp]
$%
\begin{array}{c}
\epsfxsize=7cm \epsffile{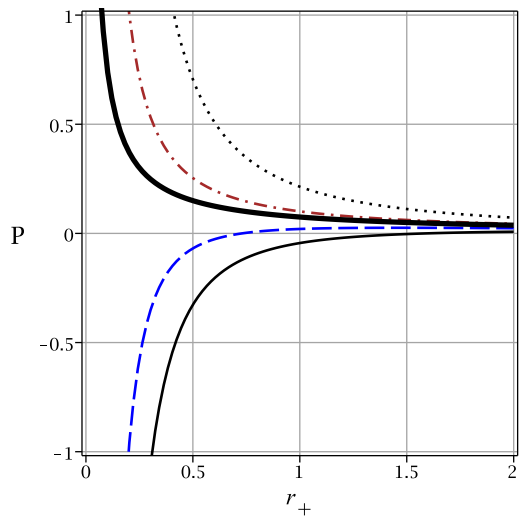}%
\end{array}
$%
\caption{$P$ versus $r_{+}$ for $d=5$, $T=0.1$ and $\Phi _{E}=0.1$; $k=1$, $%
\Phi _{M}=0$ (continuous line), $\Phi _{M}=2$ (dashed line), $\Phi _{M}=3$
(dashed-dotted line) and $\Phi _{M}=4$ (dotted line). \textbf{The bold line:}
$k=\frac{2\left( d-3\right) \left( \Phi _{M}^{2}+\Phi _{E}^{2}\right) }{%
\protect\pi ^{2}\left( d-2\right) }$ and $\Phi _{M}=4$.}
\label{FigEM2}
\end{figure}

Evidently, this relation does not explicitly depend on the horizon radius.
Remembering that horizon radius is directly related to volume of the black
hole, one can conclude the absence of critical volume. This indicates that
although the black holes with constant magnetic and electric fields could
have critical behavior, but this critical behavior is not van der Waals
like. In plotted diagrams for the pressure (see Fig. \ref{FigEM2}), one can
see that for the cases $\frac{2\left( d-3\right) \left( \Phi _{M}^{2}+\Phi
_{E}^{2}\right) }{\pi ^{2}\left( d-2\right) }\geq k$, the pressure will be a
decreasing function of horizon radius without any root or extremum. On the
other hand, for $\frac{2\left( d-3\right) \left( \Phi _{M}^{2}+\Phi
_{E}^{2}\right) }{\pi ^{2}\left( d-2\right) }<k$, pressure will acquire a
root and a maximum which is located after the root. The root and maximum are
decreasing functions of the magnetic potential. Before the root, pressure is
negative valued and at the maximum, a phase transition will take place.

It is a matter of calculation to obtain the free energy of this case in
following form

\begin{equation}
F=\frac{r_{+}^{d-3}}{16}k-\frac{\pi r_{+}^{d-1}}{\left( d-1\right) \left(
d-2\right) }P+\frac{\left( 2d-5\right) \left( d-3\right) r_{+}^{d-3}}{8\pi
^{2}\left( d-2\right) }\Phi _{M}^{2}+\frac{2\left( 2d-5\right) \pi -\left(
d-2\right) }{16\pi ^{3}\left( d-2\right) }\left( d-3\right) r_{+}^{d-3}\Phi
_{E}^{2},
\end{equation}%
which has the following root

\begin{equation}
r_{+}(F=0)=\left( 4\pi ^{2}\sqrt{\frac{P}{\left( d-1\right) \left[ \pi
^{3}\left( d-2\right) k+2\pi \left( d-3\right) \left( 2d-5\right) \Phi
_{M}^{2}+\left( d-3\right) \left( 2\pi \left( 2d-5\right) -d+2\right) \Phi
_{E}^{2}\right] }}\right) ^{-1}.
\end{equation}

The existence of root for free energy depends on the following condition

\begin{equation}
\pi ^{3}\left( d-2\right) k+2\pi \left( d-3\right) \left( 2d-5\right) \Phi
_{M}^{2}+\left( d-3\right) \left( 2\pi \left( 2d-5\right) -d+2\right) \Phi
_{E}^{2}>0,
\end{equation}

The root of free energy for this case is an increasing function of
topological factor, electric and magnetic potentials, whereas, it is a
decreasing function of the pressure.

Finally, the heat capacity for this case is obtained as

\begin{equation}
C=\frac{\left( d-2\right) \left( d-3\right) \pi ^{2}k+16\pi
^{3}r_{+}^{2}P-2\left( d-3\right) ^{2}\left( \Phi _{M}^{2}+\Phi
_{E}^{2}\right) }{64\pi ^{3}r_{+}^{2}P-4\pi ^{2}\left( d-2\right) \left(
d-3\right) k+8\left( d-3\right) ^{2}\left( \Phi _{M}^{2}+\Phi
_{E}^{2}\right) }\pi \left( d-2\right) r_{+}^{d-2}.
\end{equation}

The divergent point of the heat capacity is obtained in the following form

\begin{equation}
r_{+}(C\longrightarrow \infty )=\frac{1}{4P^{\frac{1}{2}}\pi ^{\frac{3}{2}}}%
\sqrt{\left( d-3\right) \left[ \pi ^{2}\left( d-2\right) k-2\left(
d-3\right) \left( \Phi _{M}^{2}+\Phi _{E}^{2}\right) \right] }.
\end{equation}

Considering that pressure, electric and magnetic potentials are positive
valued, the existence of divergency, hence phase transition, depends on the
satisfaction of the following condition

\begin{equation}
k>\frac{2\left( d-3\right) \left( \Phi _{M}^{2}+\Phi _{E}^{2}\right) }{\pi
^{2}\left( d-2\right) }.
\end{equation}

Comparing this condition for the presence of divergency in heat capacity
with the one for existence of root for temperature, one can see that the
following statement holds: i) in the case where the condition for having
root for the temperature holds, the condition for having divergency in heat
capacity is violated. Therefore, the temperature and heat capacity have the
same root beyond which both temperature and heat capacity are positive
valued and solutions are physical and thermally stable. ii) in the case
where the condition for having divergency in heat capacity is satisfied, the
condition for temperature having root is violated. Therefore, solutions will
have positive temperature everywhere. But before divergency, the heat
capacity is negative while after that, it will be positive. Therefore, for
this case, there is a phase transition between smaller unstable black holes
with larger stable ones. iii) finally, for the case $k=\frac{2\left(
d-3\right) \left( \Phi _{M}^{2}+\Phi _{E}^{2}\right) }{\pi ^{2}\left(
d-2\right) }$, no divergencies and roots are available for both heat
capacity and temperature. In this case, black holes are every where physical
and thermally stable. It is worthwhile to mention that in this case, the
effects of topological structure of the black holes are canceled by the
effects of the electric and magnetic potentials. It should be pointed out
that such a case only takes place for spherical black holes. In order to
show the mentioned behaviors for heat capacity, see the right panel of Fig. %
\ref{FigEM1}. Evidently, the root of temperature (heat capacity) is an
increasing function of the magnetic potential while the divergency of the
heat capacity is a decreasing function of it.

The high energy limit and asymptotic behavior of the heat capacity for $5$%
-dimensional case are in the following forms

\begin{eqnarray}
\lim_{r_{+}\rightarrow 0}C &=&-\frac{3\pi }{4}r_{+}^{3}-\frac{12P\pi ^{4}}{%
3k\pi ^{2}-4\left( \Phi _{M}^{2}+\Phi _{E}^{2}\right) }r_{+}^{5}+O\left(
r_{+}^{9}\right) , \\
&&  \notag \\
\lim_{r_{+}\rightarrow \infty }C &=&\frac{3\pi }{4}r_{+}^{3}+\frac{3}{16}%
\frac{3k\pi ^{2}-4\left( \Phi _{M}^{2}+\Phi _{E}^{2}\right) }{\pi ^{2}P}%
r_{+}+O\left( \frac{1}{r_{+}}\right) .
\end{eqnarray}

In this case, similar to previous cases, the dominant terms for high energy
limit and asymptotic behavior of the heat capacity only include horizon
radius which originated purely from gravitational part of the action. On the
other hand, the second leading term in the high energy limit includes all
the quantities of the black hole: the pressure is present in the numerator
of this term while the topological factor, electric and magnetic potentials
are in numerator. The second leading term is an increasing function of the
topological factor and pressure while it is a decreasing function of the
electric and magnetic potentials. In other words, since we regard constant
electric and magnetic fields, one expects to keep their effects
asymptotically.


\section{Conclusion}

The paper at hand investigated the higher dimensional electric-magnetic
black holes through a novel proposal. The novel proposal employed for
constructing electromagnetic tensor, resulted into components which have
valid physical properties and for Maxwell invariant, they were
distinguishable. Interestingly, the obtained metric function for this case
has electric and magnetic components of the same order of magnitude with
same factors. The event horizon and geometrical properties of the solutions
were investigated and it was shown that the number and location of root of
metric function depend on magnetic charge.

Next, thermodynamical properties of the solutions were investigated in
detail for different cases. In order to enrich the study, the concept of
extended phase space (equivalency between pressure and negative cosmological
constant) was employed as well. The investigation was done in the context of
four distinctive cases:

I) General case, in which it was shown that high energy limit of different
thermodynamical quantities except for heat capacity are governed by both the
magnetic and electric charges. The existence of the van der Waals like
behavior was reported for this case and dependency of phase transition
points on the magnetic charge was pointed out. The presence of electric and
magnetic charges were observed in denominator of the second dominant term of
the high energy limit.

II) Constant electric field in which comparing to previous case, the high
energy limit of the thermodynamical quantities, except heat capacity, was
determined by magnetic charge. The presence of electric potential was
evident in the second dominant term in the numerator, even for heat
capacity. Interestingly, it was possible to trace out the effects of
topological structure of the black holes by suitable choices of the electric
field. It was shown also, that in the absence of magnetic charge, no van der
Waals like phase transition is present, although there exists a phase
transition. The type of phase transition was different for this case.

III) Constant magnetic field. For this case, the dominant term of high
energy limit of different thermodynamic quantities, except heat capacity,
included only electric charge while the effects of magnetic potential were
observed at the second leading term. Contrary to the previous case, here, it
was possible to cancel out the effects of topological parameter of black
holes via a specific condition involving the magnetic potential. The
existence of van der Waals behavior and dependency on number of the
divergencies in heat capacity and stability condition on magnetic potential
were highlighted. It was shown that for super-magnetized black holes (large
magnetic field), thermal structure of the black holes are modified on level
of the absence of van der Waals like behavior and phase transition. In other
words, for super magnetized black holes, heat capacity is a smooth function
of the horizon radius.

IV) Constant magnetic and electric fields, simultaneously. In this
particular case, there was no van der Waals like behavior available for
black holes under any circumstances. Although the heat capacity enjoyed a
divergency, hence critical behavior, in this case for specific values of the
magnetic potential, the type of phase transition was different from van der
Waals like phase transition. It was also shown that for this case, it is
possible to remove the effects of topological parameter of the black holes
alongside with electric and magnetic potentials for specific choices of
these parameters. In that case, both the high energy limit and asymptotic
behavior of black holes are governed only by pressure, hence cosmological
constant.

One of the main goals here was to provide the possibility of investigating
dyonic (electric-magnetic) black holes in higher dimensions. The novel
proposal here provided such a possibility. The next step would be
understanding the holographical aspects of this proposal in higher
dimensional solutions and see how higher dimensionality would modify the
feromagetism/diamagnetism phase transitions. In addition, higher curvature
gravities that are dimension-dependent could be coupled with this theory, as
well, which help understanding the effects of these higher derivative
gravity theories on the feromagetism/diamagnetism phase transitions and
holographical principles. Furthermore, the investigation of
superconductivity in the presence of this magnetic field is yet another
interesting subject which we leave for a future work.

\begin{acknowledgements}
We thank both Shiraz University and Shahid Beheshti University
Research Councils. This work has been supported financially partly
by the Research Institute for Astronomy and Astrophysics of
Maragha, Iran.
\end{acknowledgements}

\end{document}